\documentclass{article}
\usepackage{graphicx}
\usepackage{amssymb,amsmath,amsfonts}
\usepackage{colortbl}
\usepackage{pstricks}
\usepackage{geometry}
\usepackage[english]{babel}
\usepackage[latin1]{inputenc}
\usepackage{latexsym}

\newcommand{\field}[1]{\mathbb{#1}}
\title{\textbf{ Computing transition rates for the 1-D stochastic Ginzburg--Landau--Allen--Cahn equation for finite-amplitude noise with a rare event algorithm}}
\author{Joran Rolland\footnote{Institut Non lin\'eaire de Nice, UMR 7335, CNRS
              1361, route des Lucioles, 06560 Valbonne, France
  \emph{Present address:}
              Institut f\"ur Atmosph\"are und Umwelt, Goethe Universit\"at
              Altenh\"oferallee 1, 60438 Frankfurt, Deutschland
              {\sc rolland@iau.uni-frankfurt.de}    }~, Freddy Bouchet\footnote{
              Ecole Normale Sup\'erieure Lyon, CNRS
              46, all\'ee d'Italie (Site Monod) 69364 Lyon Cedex 07, France
              {\sc Freddy.Bouchet@ens-lyon.fr}}~, Eric Simonnet\footnote{
              Institut Non lin\'eaire de Nice, UMR 7335, CNRS
              1361, route des Lucioles, 06560 Valbonne, France
              {\sc eric.simonnet@inln.cnrs.fr}}}

\date{\today}
\begin{document}

\maketitle

\begin{abstract}
In this article we compute and analyse the transition rates and duration of reactive trajectories of the stochastic 1-D Allen-Cahn equations
for both the Freidlin-Wentzell regime (weak noise or temperature limit) and finite-amplitude white noise, as well as for small and large domain. We demonstrate that
extremely rare {\it reactive trajectories} corresponding to direct transitions between two metastable states are efficiently computed using
an algorithm called {\it adaptive multilevel splitting}. This algorithm is dedicated to the computation of rare events
and is able to provide ensembles of reactive trajectories in a very efficient way.
In the small noise limit, our numerical results are in agreement with large-deviation
predictions such as instanton-like solutions, mean first passages and escape probabilities.
We show that the duration of reactive trajectories follows a Gumbel distribution like for one degree of freedom
systems. Moreover, the mean duration growths logarithmically with the inverse temperature.
The prefactor given by the potential curvature grows exponentially with size.
The main novelty of our work is that we also perform an analysis of reactive trajectories for large noises and large domains. In this case, we show
that the position
of the reactive front is essentially a random walk. This time, the mean duration grows linearly with the
inverse temperature and quadratically with the size. Using a phenomenological description of the system, we are able to calculate the transition rate, although the dynamics is described by neither Freidlin--Wentzell or Eyring--Kramers type of results. Numerical results confirm our analysis.
\end{abstract}
\begin{flushleft}
\end{flushleft}

\section{Introduction}
\label{intro}
\begin{figure}
\centerline{
\includegraphics[width=10cm]{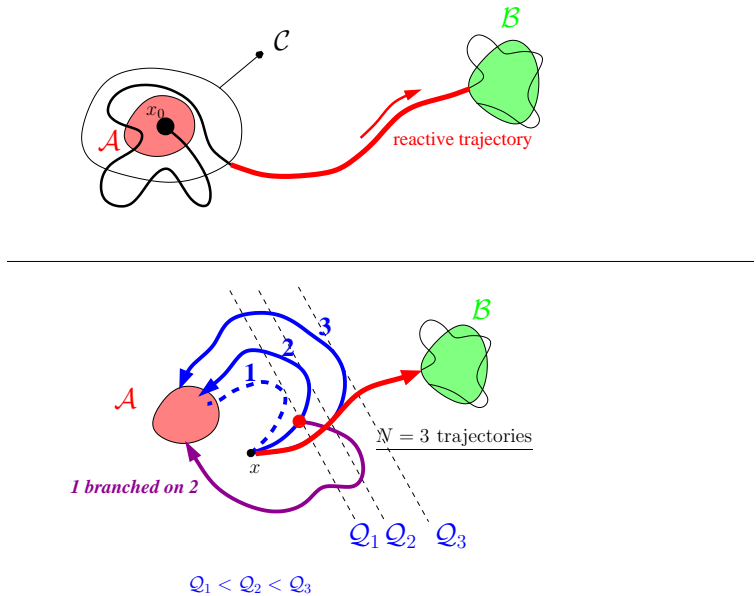}
}
\caption{Upper panel: a first passage trajectory (with mean duration called $T$ in the text)
is shown in black starting from a fixed point $x_0$ and its final portion
corresponding to the so-called reactive trajectory (with mean duration called $\tau$ in the text)
is shown in red.
Lower panel: after generating $N$ i.i.d. trajectories starting from $x$, AMS is eliminating the
trajectory with the worst excursion (here trajectory 1 corresponding to ${\cal Q}_1$)
with respect to the reaction coordinate $\Phi$ (see Appendix)
and replaces it using information picked from a randomly chosen trajectories among the $N-1$ ``better" ones (here
$N=3$ and trajectory 2 is chosen by 50$\%$ chance).
This is done by restarting a
new trajectory from the initial condition (the red dot in the figure).
This process is repeated until all trajectories has reach the set $B$ before $A$.}
\label{sketch}
\end{figure}

This article is a numerical and theoretical analysis of {\it reactive trajectories} between two metastable states of the 1-D stochastic Allen-Cahn equation.
Reactive trajectories are paths which directly connect two stable or metastable sets $A$ and $B$
(see Fig. \ref{sketch}).
Metastability is ubiquitous in physics and chemistry, for instance in solid state physics \cite{bray}, physical chemistry \cite{jpcm} or
molecular dynamics \cite{Mol}. It has also been studied recently in new fields, for instance in fluid dynamics systems, like in
2-D Navier--Stokes equations
\cite{sns}, turbulent Von-Karman flows \cite{Vonkarman}, turbulent convection \cite{convection},
or wall turbulence \cite{couette}. In oceanic and atmospheric flows, well-known examples are the
bimodal behavior of the Kuroshio paths
\cite{kuro}, thermohaline circulation in the North-Atlantic \cite{Ey,cessi} or zonal-blocking transitions in the midlatitude atmosphere \cite{PV} ; in geophysics, one can also think of reversal of dynamos \cite{dyn}.
Many of these systems can be modeled by
stochastic partial differential equations (SPDEs).

In the last three decades, the stochastic Allen--Cahn equations and more generally reaction-diffusion
equations have become central in the mathematical study of SPDEs
\cite{faris,dptb,bbm1,fb,mkpde,mh}. They exhibit both relative simplicity and nontrivial behavior
such as bimodality and effects of large dimensions. They are widely used for
the study of phase-ordering kinetic (see e.g. \cite{bray,hh77}). In particular,
the stochastic Allen--Cahn equations can be found in fluid mechanics as a convenient model of long-wavelength
modulations in turbulent Couette flows (where it is termed Ginzburg--Landau equation) \cite{couette}, or in
the context of quantum mechanics as a model of a double-well oscillator (see \cite{faris}). In these systems, the metastable state comprise of a single phase in the whole domain. Meanwhile, during reactive trajectories the domain contains subdomains of different phases separated by \emph{fronts}.

For systems with one degree of freedom like stochastic differential equations (SDEs),
solutions of the Fokker--Planck equations can often be obtained and
analyzed analytically \cite{tpt,ht,ons}.
As a consequence, all the statistics of reactive trajectories can be inferred.
However, when the system of interest has too many degrees of freedom, an explicit use
of such methods is often impossible. In such situations, several approaches can bring
insight on the transitions.
The discussion mostly depends on the noise amplitude.
For relatively large noise, transitions between metastable states can be
computed by direct numerical simulations. As the noise intensity decreases,
these transitions occur with smaller transition rates and are therefore more difficult
to observe using direct numerical simulations. In such a case, a possibility is to consider the small-noise limit theories
such as the Freidlin--Wentzell \cite{FW} or  the Eyring--Kramers \cite{ha,Kramers,Eyring,JSL} asymptotic theories for transition rates.
The Freidlin--Wentzell large-deviation theory describes the most probable
transition called {\it instanton} and the Eyring--Kramers expression gives the full formula, with the sub-exponential prefactor, for the mean first transition time or equivalently the transition rate. This formula holds strictly for gradient stochastic differential equations, but can be generalised (with modification) to a larger class of non-gradient system \cite{rey}.
The case of SPDEs offers another level of difficulty: large-deviations results are very
difficult to establish in general. For instance, it is not even known
if a large-deviation principle exists for the Allen--Cahn equation in large dimension of space
\cite{mh}. However, since the seminal work of Faris \& Jona-Lasinio \cite{faris}, the
1-D Allen--Cahn equation appears to be now well understood (see e.g. \cite{bbm1,fb,mkpde}), and follow the classical finite dimension phenomenology.
Note that even for finite-dimensional systems, serious theoretical difficulties
may also arise when the saddle points of the deterministic dynamics are degenerated \cite{nq1},
or even worst, when the local attractors representing the metastable states are
non-isolated \cite{bt_niso}.
Nevertheless, these large-deviation theories have been used with success on some
SPDEs (see e.g. \cite{E_KS}) including the stochastic Allen--Cahn equations
where finite-time instantons have also been numerically computed with great accuracy \cite{VE}. Another relevant approach of SPDEs is the direct
computation of instantons \cite{gr1,gr2} using essentially a geometrical approach \cite{gr3}.

In the 1-D Allen--Cahn case, if the domain is not too small, instantons and transitions between metastable states take the form of fronts
propagating from one side of the domain to the other. Previous results like the calculation of E \emph{et al.} \cite{VE} as well
as the small-noise limit theories in general do not take into account finite-amplitude noise effects. As will be discussed in this article, the range of validity of the Freidlin--Wentzell asymptotic regime is however extremely limited. The regime with noise too large to reach the Freidlin-Wentzell asymptotic, but still small enough for the event to be rare, occupies a very large area of the parameter space for partial differential equations. For symmetric potentials, a theoretical analysis of metastability for the 1-D Allen--Cahn case can be tackled if the size of the domain is large compared to the front thickness \cite{wall_pre,BBMP,kohn,fun1,fun2}. In this case, the
propagation of the fronts is a random walk. All things considered, the approaches are
very diverse and leave whole areas of the parameter space unexplored. In particular, the limit
of validity of Eyring--Kramers and Freidlin--Wentzell theories has never been considered precisely. We propose here to give a global description of the statistical behavior of 1-D Allen--Cahn equation for
arbitrary noise amplitude as well as finite-size domains.

In order to do that, we consider a new numerical tool dedicated to the computation of rare events
and in particular reactive trajectories. This algorithm  is called
{\it adaptive multilevel splitting} (AMS)
and has been proposed rather recently by C\'erou \& Guyader \cite{cg07}. It corresponds in fact to a modern adaptive
 version of the classical multilevel splitting algorithm introduced in the 50s by Kahn \& Harris \cite{KH} and Rosenbluth \& Rosenbluth \cite{RR}.
Classical splitting algorithms have been very popular in the last few decades and have
been applied in many areas of physics (chemistry, molecular dynamics and networks).
This class of algorithms is now better understood from
a mathematical point of view and have become an important area of research
in probability theory \cite{PDM}. AMS appears to be very efficient and robust, it is moreover much
less sensitive to tuning parameters than the classical versions and can in principle be applied
to systems with a large number of degrees of freedom. In fact, the main constraint is to
work with Markovian systems.
The idea of the algorithm is rather simple
to grasp: it is essentially a go-with-the-winner strategy
(or successive iterations of the weakest link game) in the space of trajectories.
At the start of the algorithm, one generates an ensemble of trajectories which are
statistically independent and identically distributed.
By successively eliminating the bad trajectories and generating new ones by splitting them
from the good trajectories, one is able to obtain easily a whole set of trajectories connecting
one metastable state to another. In principle, this can be done for any amplitude
of the noise and possibly for more exotic colored noise as well. One is then able to perform
 statistics for rare events such as estimating the distribution of the duration of reactive
trajectories. Although theoretical subtleties still remain regarding convergence
properties \cite{STCO,poi,jcp,pdm2},
AMS approach is likely to offer one of the most efficient tool for handling rare events computation.
Until now this algorithm has only been applied to SDEs  with a small number of degrees of freedom, and
mostly for test and convergence analysis. One of the main aim of this study is demonstrate the practicality and usefulness of this algorithm in high dimensional and spatially extended systems as well. Moreover, as will be shown, the numerical results will provide a much broader view on the theory of reactive trajectories. A sketch of the algorithm is shown in Fig.\ref{sketch}.
A detailed discussion is also given in the Appendices~\ref{Ad} and~\ref{Ac} together with a pseudo-code description.

This article shows several results. We are able to obtain a
complete theoretical and numerical
statistical description of the reactive trajectories for the 1-D stochastic Allen--Cahn equation
in the parameter plane spanned by the inverse temperature $\beta$ and the length $L$ of the domain.
Those are the only control parameters of the Allen--Cahn equation once properly rescaled.
We also discuss the type of reactive trajectories, the transition rate, the mean first passage time and the
duration of reactive trajectories as a function of $\beta$ and $L$.
We show that only a small portion of the parameter plane for large $\beta$ and
 small $L$ leads to ensemble of reactive trajectories that concentrate close to the most probable one, as could be expected from Freidlin--Wentzell theory and instanton phenomenology. We investigate the transition rate and the mean first-passage time and recover the classical
Eyring--Kramer law in this narrow band of the parameter plane. We show in particular that the prefactor has an exponential dependence in the size. We are also
able to distinguish the limit of validity of this law.
In some parts of the parameter space, for finite-amplitude noise, reactive trajectories
are close to instantons from a geometric point of vue, but their temporal evolution is not close to the instanton one, by contrast to the Freidlin--Wentzell regime case. In particular, they are characterized by a random-walk propagation of their fronts. For a very large size, the system presents the properties of a potential with a extremely flat saddle. This may be related to the case of non-quadratic saddles for which classical results breakdown\cite{nq1,VEL}. Note that this can also be studied in a case where the global minima of the potential do not have the same height \cite{VEL}. The dynamics is described by an effective potential whose curvature at the saddle in exponentially small in $L$. We can write and solve a coupled master/Fokker--Plank equations that describes the position of the front. From this, we can calculate the transition rate and show that its prefactor is not exponential in $L$, as would be expected from Eyring-Kramers expression with an exponentially small curvature, but rather polynomial. However, the transition rate retains its large deviations properties in $\beta$. This analytical result is verified by numerical computations using AMS. We are also able to investigate the statistics of the reactive trajectory durations as well.
We find two distinct regimes of large curvature and small curvature at the saddle where the duration
scales like ${\rm exp}(L/\sqrt{2}) \ln \beta$ and $L^2 \beta$ respectively. Moreover, in the large
curvature regime, the distribution of durations follows a Gumbel distribution like for
one degree of freedom systems \cite{asy}. As far as we know, these are new results.

The structure of this work is as follow. We first introduce the 1-D stochastic Allen--Cahn and its
deterministic solutions in Section \ref{S2}. We then perform in Section \ref{S3}
a theoretical analysis of metastability in Allen--Cahn equation using small noise theories such
as Freidlin--Wentzell large deviation theory and Eyring--Kramers law.
Section \ref{S4} extensively makes use of the AMS algorithm for computing numerically ensemble
of reactive trajectories in various regimes of the Allen--Cahn equation.
We will be particularly interested in the comparison between theory and numerics in
the range of parameters where both are valid.
We discuss these results in the conclusion (\S~\ref{S5}).

\section{The stochastic Allen--Cahn equation}
\label{S2}
We present here the 1-D stochastic Allen--Cahn equation with a definition of the control parameters and
some properties of the stationary solutions of the associated deterministic Allen--Cahn equation.

\subsection{Introduction}
\label{S2.1}
The nondimensional 1-D stochastic Allen--Cahn equation with
Dirichlet boundary conditions reads
\begin{equation}\label{gl1}
\partial_t A = \partial_x^2 A + (A-A^3) + \sqrt{\frac{2}{\beta}} \eta, A(0) = A(L) = 0,
\end{equation}
where $\eta$ is a white noise in space and time,
$<\eta(x,t)\eta(x',t')> = \delta(t-t')\delta(x-x')$ \cite{gar,VK}. It has several names:
it is often called the Allen--Cahn equation in mathematics and solid state physics \cite{bray} and the
Chaffee--Infante in climate science. In non-linear physics, it is one of the several instances of Ginzburg--Landau equation \cite{CH}. In the study of the deterministic front propagation in biology, it is a particular case of the Fisher--Kolmogorov equation \cite{BD}. After change of temporal, spatial and amplitude scales, there are only two free parameters, the nondimensional
inverse temperature $\beta$ and the nondimensional size of the domain $L$. With this scaling choice, the size of the boundary layers,
fronts and coherent length is of order one.
We recall an important property of the Allen--Cahn equation
which is essential for theoretical considerations: it is a gradient system (for the $\mathbb{L}^2$ scalar product) with respect to the potential $V$
\begin{equation}\label{gl2}
\partial_t A = -\frac{\delta V}{\delta A} + \sqrt{\frac{2}{\beta}} \eta,~
V = \frac12 \int_0^L \left( - A^2 + \frac{A^4}{2} + (\partial_x A)^2 \right) dx.
\end{equation}
This formulation allows one to have an easier understanding of
the phase space structure and dynamics of the Allen--Cahn equation.

\subsection{Stationary solutions}
\label{S2.2}
\begin{figure}
\centerline{\includegraphics[width=7cm]{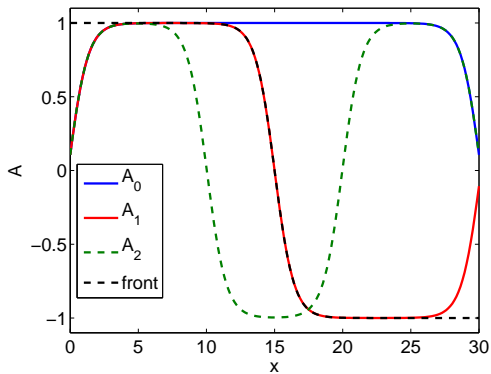}}
\caption{Example of stationary solutions of the 1-D deterministic Allen--Cahn equation
in a domain of size $L=30$. The solution $A_1$ with one front is compared to
the analytical solution of the front Eq.~(\ref{front}). Equation (\ref{pot}) is solved by finite-difference using a Newton
method with initial condition $\sin \pi(n+1)x$.}
\label{figsols}
\end{figure}

Stationary solutions of the deterministic Allen--Cahn equation solve the Sturm--Liouville problem (see Faris \& Jona-Lasinio \cite{faris}
 \S~7 and references therein)
\begin{equation}\label{pot}
\partial_x^2 A = -\frac{dU}{dA},~A(0)=A(L) = 0,
\end{equation}
where $U = A^2/2 - A^4/4$. Depending on $L$, several solutions may exist, the stable
ones play the role of the metastable states in the stochastic Allen--Cahn equations whereas saddle points indicate possible pathways between them.
Equation (\ref{pot}) can be solved using the analogy of a pendulum of period $L$ in a potential $U$ where $x$ plays the role
of time. Provided $L$ is not too small,
there are two stable solutions denoted $A_0^\pm$ and one unstable solution $A_0$ (Fig.~\ref{figsols})
\begin{equation}
A_0 =0, A_0^\pm = \pm 1 ~+ \mbox{``boundary-layer corrections"}.
\end{equation}
The two stable solutions correspond to the
global minima of $V$ (Eq.~(\ref{gl2})).  Due to the existence of Dirichlet boundary  conditions
at $x=0,L$, boundary-layer for $A_0^\pm$ should also be considered.
As $L$ is increased, pairs of unstable solutions denoted $(-A_n,A_n)$ successively appear.
It has been demonstrated in \cite{faris} that the critical values of $L$ for which this happens are
\begin{equation}\label{cond}
L = (n+1) \pi
\end{equation}
These solutions have exactly $n$ fronts (called kinks in field theory \cite{bray}) and
$n$ zeros in the domain $]0,L[$. Using the dynamical analogy of the pendulum, it
means that the solution are periodic with period $2L/(n+1)$. Note that the Allen--Cahn equation without boundary conditions
and the diffusive part $\partial^2_x A$ has a whole family of stationary solutions solving $A^2=1$.
These solutions can arbitrarily jump from the values +1 and -1 at any points of the domain.
The diffusive term together with the boundary conditions select some
of these solutions.

Figure \ref{figsols} shows three of the solutions of (\ref{pot}) for $L=30$ where $1 + 2 \times 8$ solutions coexists due to the
symmetry $A \to -A$ (see (\ref{cond})).
The figure shows the stable solution $A_0^+$ and two saddle solutions $A_1$ (one front) and $A_2$ (two fronts).
The size of the fronts and boundary layers is $O(1)$ as expected and the solutions take values close to $\pm 1$ away from
the boundary layers.

It is possible to give an analytical expression of the saddle $A_1$ locally near $x = L/2$ for large $L$.
We denote $A_1^\infty$ the solution which corresponds to a solution of $\partial_x^2A=-dU/dA$ on $\mathbb{R}$ with limit conditions
$\lim_{x \to \pm \infty} A_1(x) = \pm 1$. It takes the exact form
\begin{equation}\label{front}
A_1^\infty(x) = \tanh \left( \pm \frac{x}{\sqrt{2}} \right).
\end{equation}
This is indeed the classical  kink formula \cite{bray,wall_pre}.  The term $\pm$ comes from the symmetry of the problem $A \to -A$,
it is $+1$ for an ascending front or kink and $-1$ for a descending front or anti-kink. A good numerical match is found in Fig. \ref{figsols}
between this expression and the exact solution $A_1$ away from the external boundary layers.
Note that for more complex potentials, Taylor expansion of the potential would yield local expressions as well (see  \cite{cessi} \S~B).

  The analysis presented by Faris \& Jona-Lasinio \cite{faris} indicates how many fixed point the deterministic system has and whether they are stable. In the case of Dirichlet boundary conditions, it does not give the eigenvalues of the Hessian of $V$ at said fixed points: these eigenvalues indicates how stable or unstable the fixed points are. Using analytical description of the type of Eq.~(\ref{front}) we are able to give an approximation of the one negative eigenvalue of the Hessian of $V$ at $A_1$ (the most important unstable fixed point for reactive trajectories). We describe the result in section~\ref{exponential_interactions} and detail the derivation in appendix~\ref{Ap_exponential_interactions}. In particular, we are able to show that this eigenvalue converges toward $0$ exponentially with a rate $\exp(-L/\sqrt{2})$. This indicates that the potential $V$ becomes extremely flat at the saddle $A_1$. This flatness is the main cause of the diffusive regime described in section~\ref{S3.3}.

\section{Theoretical methods for metastable systems}
\label{S3}
We present here the theoretical approaches we use to analyse metastability. We also complement these by
specific discussions and results whenever it is necessary.
First, we discuss the Freidlin--Wentzell large-deviation approach in subsection \ref{S3.2}, then Eyring--Kramers formula and the distribution of reactive trajectory durations in subsection \ref{S3.1},
then the front dynamics description in subsection \ref{S3.3} and summarize these results in a phase diagram
in subsection \ref{S3.4}.

\subsection{Large deviations, Freidlin-Wentzell theory, and Instantons}
\label{S3.2}
We consider a gradient dynamics in a potential $V$ with a weak noise of amplitude $\sqrt{2/ \beta}$. We consider the transition from the basin of attraction of one local minima $u_0$ of $V$ to another one $u_1$. We assume that the two basins of attraction touch each other at at least one saddle point. Generically the transition rate $\mu$ from the basin of attraction $u_0$ to the one of $u_1$ follows a large deviation result
\begin{equation}
\underset{\beta \rightarrow \infty}{\lim}  -\frac{\ln \mu}{\beta} = V(u_s)-V(u_0),
\label{large-deviation-rate}
\end{equation}
where $V(u_s)$ is the minimum of $V$ at any saddle points connecting the two basin of attraction of $u_0$ and $u_1$. When the lowest saddle is unique, the reactive trajectories concentrate close to a single path called instanton--anti-instanton. This phenomenology can be proven within the Freidlin-Wentzell theory \cite{FW}. As far as the 1D-Allen--Cahn equation is concerned, the proof of large deviation principles has first been established by Faris \& Jona-Lasinio \cite{faris}. In this section, we explain heuristically those results, and describe the classical results about the saddle points and instantons of the 1D-Allen--Cahn equation.

We consider a partial differential equation
\begin{equation}
du = F(u)dt + \sqrt{\frac{2}{\beta}} dW_t,
\label{SPDE}
\end{equation}
where $u(t) \equiv u({\bf x},t)$, ${\bf x} \in \field{R}^d$, $u(t=0)=u_0({\bf x})$ is an initial condition
and $W_t$ is a (space-time) Wiener process. We call ${\cal V}={\cal V}(u_0,u_1,\tau)$ the set of trajectories which start from $u_0$
and end at $u_1$ at time $\tau$. A path integral approach (Initiated by Onsager and Machlup \cite{OM}) allows to formally express the transition probability from the state $u_0$ at time $t=0$ to the state $u_1$ after time $t=\tau$ as
\begin{equation}
P(u_1,\tau;u_0,0) =  \int_{\cal V} {\rm exp}
\left(-\frac{\beta}{2} \mathcal{L}[u] \right)
{\cal D}[u],
\label{Onsager-Machlup}
\end{equation}
where {\it the action} is
$$
\mathcal{L}[u,\tau] = \frac12 \int_0^\tau \left|\frac{du}{dt} - F(u)\right|^2 dt.
$$
In the low noise limit, the transition rate $\mu$ between attractors $u_0$ and $u_1$ is defined as $P(u_1,\tau;u_0,0) \underset{1 \ll \tau \ll 1/\mu}{\sim}\mu \tau$. From the path integral (\ref{Onsager-Machlup}), taking first the limit $\beta$ to infinity through a saddle-point approximation, and then the limit $\tau\rightarrow \infty$  yields the large-deviation result
\begin{equation}
\lim_{\beta \to \infty} -\frac{1}{\beta} \ln P = \frac{1}{2} \inf_{u\in \tilde{{\cal V}} } \mathcal{L}[u,\infty]\,,\, \tilde{{\cal V}}=\{ u(t) \backslash \lim_{t\rightarrow -\infty}u(t)=u_0\,,\, u(0)=u_s \}
\label{rate_action}
\end{equation}
where $u_s$ is a saddle point at the boundary of the $u_0$ basin of attraction.
Remarkably, Freidlin and Wentzell have shown that this saddle approximation is indeed valid
provided $u_0$ is a fixed isolated attracting point of the deterministic system. Moreover, for gradient systems where
$F(u) = -\frac{\delta V}{\delta u}$ the action minimization is easy. For any $u_i$ in the basin of attraction of $u_0$, we call a relaxation path from $u_i$ to $u_0$ a solution of
\begin{equation}
\label{flurel}
\frac{du}{dt} = - \frac{\delta V}{\delta u}\end{equation}
with initial condition $u(t=0)=u_i$. We call a fluctuation path a time reversed relaxation path \cite{FB_JL}. Then the unique minimizer of the infinite time action from $u_i$ to $u_0$ is a relaxation path, the action minima is then equal to zero ; and the unique minimizer of the infinite time action from $u_0$ to $u_i$ is a fluctuation path, and the action minima is then equal to $2V(u_i)-2V(u_0)$ \cite{FW}. We call an instanton a solution of (\ref{flurel}) with $\lim_{t \rightarrow -\infty} u(t)=u_0$ and $\lim_{t \rightarrow \infty} u(t)=u_s$ (the instanton is then defined up to an arbitrary time shift). An anti-instanton is a time reversed instanton. Then it is easy to understand that  $\inf_{u \in {\cal V}} \mathcal{L}[u,\infty] = 2V(u_s)-2V(u_0)$, and that formula (\ref{rate_action}) leads to the result (\ref{large-deviation-rate}).
 Because the saddle point $u_s$ is at the boundary of the basin of attraction of $u_0$, we stress however that there is not strictly speaking a relaxation path from $t=0$ at $u_s$ to $u_0$. As a consequence, no minimiser exist for Eq.~(\ref{rate_action}). Any path that approximates an instanton from $u_0$ to $u_s$ followed by an anti-instanton from $u_s$ to $u_1$ has an action that is close to $2V(u_s)-2V(u_0)$. We discuss this further with the Allen--Cahn example below.\\

\begin{figure}
\centerline{
\includegraphics[width=18cm]{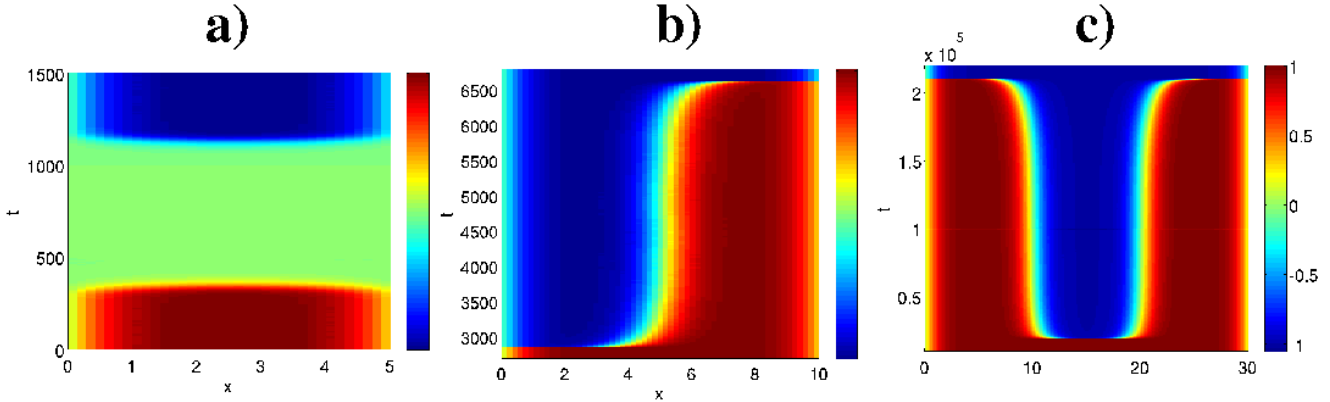}}
\caption{Space-time representations of relaxation-fluctuations paths for three different saddles:
a): $A_0=0$ $(L=6)$, b): $A_1$ $(L=10)$, c): $A_2$ $(L=30)$. Only a) and b) can be instantons.}
\label{figinst}
\end{figure}

We now apply these results to the Allen--Cahn equation using $u_0 = A_0^+$ and $u_1 = A_0^-$ and for different
saddles. For instance, for $u_s = A_0 = 0$, Fig. \ref{figinst} is a space-time representation of the succession of an approximate instanton--anti-instanton trajectory.  One obtains a global flip
of the solution. We refer such a solution as a ``flip''. We note that the duration of the trajectory close to the saddle point (the green part on Fig. \ref{figinst}) is chosen arbitrarily and that the minimum of the infinite time action is obtained in the limit where this duration is infinite. When the saddles are $A_n$ for $n \geq 1$, the instantons--anti-instantons correspond to
some fronts propagating either from the boundaries (Fig. \ref{figinst}b)
or in the bulk of the domain (Fig. \ref{figinst}c). We will term ``front" trajectories such solutions.
The sharp transition which can be seen in Fig. \ref{figinst}b,c corresponds in fact to
a timescale much smaller than the motion near the saddle and not to some numerical artifacts. For these cases too, the time spent by this approximate instanton--anti-instanton trajectory close to the saddle point is arbitrary.
From the formal interpretation of the large deviation through path integrals, as described above, we expect the distribution of reactive trajectory to concentrate close to approximate instanton--anti-instanton trajectories. For the actual dynamics with small but finite noise, the time spent close to the saddle point is however not arbitrary. This is a random variable, with an average value that scales like $\ln(\beta)$ in the limit of small noise $1/\beta$, as discussed further in section \ref{S3.3.1}.

\begin{figure}
\centerline{
\includegraphics[width=6cm]{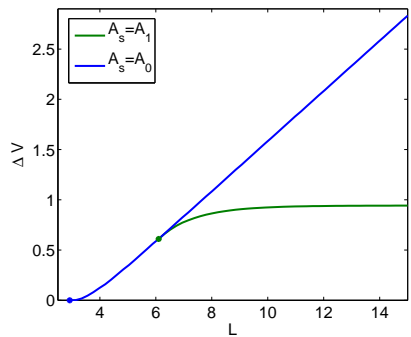}
}
\caption{Numerical computation
of the potential difference as a function of $L$ for $A_s=0$ et $A_1=0$.}
\label{deltav}
\end{figure}

From the previous discussion, as several saddles exist, several instanton--anti-instanton trajectories can join $u_0 = A_0^+$ and $u_1 = A_0^-$. In the low noise limit, the relevant one is the one that minimizes $V(u_s)-V(u_0)$.
For large $L$, the potential difference $\Delta V_n \equiv V(A_n)-V(A_0^\pm)$ for $n \geq 1$ can be investigated
analytically. The idea is simply to notice that $A_{n}$ and $A_0^\pm$ mostly
differ in the neighborhood of the fronts provided $L$ is large enough so that these fronts
are well separated. In practice, this requires that $L \gg n \sqrt{2}$. Let us look at
the case $n=1$, the solution $A_1^\infty$ (see eq.\ref{front}) is a good
approximation of the saddle so that in this case using $|A_0^\pm| \simeq 1$, one can write
$$
\Delta V_1 \simeq
\int_{-\infty}^\infty \left(\frac14 {A_1^\infty}^4 -\frac12 {A_1^\infty}^2 + \frac12(\partial_x
A_1^\infty)^2 + \frac14 \right) dx.
$$
After simple algebra, the above integral gives $\Delta V_1 \simeq \frac{2 \sqrt{2}}{3}$.
The case of several fronts follows the same line of reasoning, each front gives a similar
contribution in the potential difference provided $L$ is large and finally
\begin{equation}\label{potdiff}
\Delta V_n \simeq n \frac{2 \sqrt{2}}{3}, n \geq 1.
\end{equation}
The conclusion is that minimizers can only be instanton--anti-instantons going either through $A_0$ or $A_1$ where
the potential difference is the smallest. A more precise analysis shows that for $L<2\pi$, the saddle $A_1$ does not exist. It appears through a pitchfork bifurcation for $L=2\pi$. For $L>2\pi$, $\Delta V_1<\Delta V_0$ as illustrated on figure \ref{deltav}. We thus conclude that for $L<2\pi$ only the flip instanton exists, while for $L>2\pi$ the relevant instanton--anti-instanton is the one front transition.

\subsection{Transitions rates, mean first passage times and Eyring--Kramers'law}
\label{S3.1}
\subsubsection{General discussion}
We consider the stochastic differential equation (or partial differential equation) (\ref{SPDE}), assuming in this section that the force is gradient: $F(u)=-\nabla V(u)$. We want to compute, asymptotically for small noise ($\beta \rightarrow \infty$) the mean transition time to go from one potential minimum say $u_0$ to another $u_1$ assuming that the two basins of attraction are connected through a saddle point $u_s$. If we have two local minima only, any trajectory starting from $u_0$ will eventually hit a small neighborhood of $u_1$, after having possibly wandered around $u_0$ for a very long time. We recall that the time for a trajectory to reach $u_1$, is a random variable $\mathcal{T}$. The mean first passage time $T \equiv \field{E}^u_0\left[ \mathcal{T} \right]$ is equal to the inverse of the transition rate $T=1/\mu$, asymptotically for small noise. The trajectory between $t=0$ and $t=\mathcal{T}$ are called a first passage trajectory.

Provided the saddle $u_s$ is non degenerate, the Hessian $\nabla^2 V$ at $u_s$ has a unique negative eigenvalue $\lambda_s(u_s)$ (corresponding
to the dim-1  unstable manifold). It is a classical result that
asymptotically for small noise (see \cite{ha,Kramers,Eyring,JSL}), the mean first passage time is equal to
\begin{equation}\label{mfpt}
T = \frac{2\pi}{|\lambda_s(u_s)|}
\sqrt{\frac{|{\rm det}\nabla^2 V(u_s)|}{{\rm det}\nabla^2 V(u_0)}} {\rm exp} \left(\beta(V(u_s) - V(u_0)) \right).
\end{equation}
Geometrically, the negative eigenvalue $\lambda_s$ of the Hessian corresponds to the lowest curvature
of the potential $V$ at the saddle and controls the time scale of the dynamics around it. We note that this Eyring-Kramers formula generalize, in this gradient case, the large deviation result discussed in the previous section.
The distribution of $\mathcal{T}$ is in general exponential, i.e. is
$\Pr(\mathcal{T} \simeq s) = \frac{1}{T} {\rm exp} (-s/T)$ (see (\ref{mfpt})) \cite{VK}.

A difficulty is to give a meaning of these results in the context of SPDEs when $u$
now belongs to a functional space. For instance, it is easily seen for the 1D Allen--Cahn equation that
the product of the Hessian eigenvalues diverges. However, it can be shown that in the case of 1-D Allen--Cahn equation, the ratio
which appears in eq. (\ref{mfpt}) can be well-defined (see \cite{bbm1,fb,mkpde}).
Another level of difficulty can appear however in dimension larger than 2 at the level of
the large deviations. First, it appears that Allen--Cahn equations are in fact ill-defined when the
spatial dimension is strictly larger than one (which is not the case, we have considered in this
work) \cite{dptb,wal}.
 One has to renormalize the equation properly so that large-deviation results can then
be obtained in dimension 2 and 3 which are very difficult to demonstrate \cite{mh}. In
fact, the question is largely open in dimension larger than 3. At the level of the prefactor
of Eyring-Kramer law nothing is known even in dimension 2. Only the one-dimensional case is
well-understood \cite{faris,bbm1,mkpde}.

\subsubsection{Application to the Allen--Cahn equation}
\label{exponential_interactions}

We now discuss the behaviour of the prefactor
\begin{equation}
\frac{2\pi}{\sqrt{|\lambda_{s}(A_s)|}}\sqrt{\frac{\left|\prod_{i=2}^\infty \lambda_{i}(A_s)\right|}{\left|\prod_{i=1}^\infty \lambda_{i}(A_0)\right|}}
\label{eqpref}
\end{equation}
of the mean first passage time.  A general property of such systems is that $\lambda_s(A_s)$ goes to zero in the infinite size limit while the rest of the prefactor is unchanged. We therefore concentrate our effort on that eigenvalue $\lambda_{s}(A_s) \le 0$: it controls the size dependence of the mean first passage time.

\begin{figure}
\centerline{\includegraphics[width=5.5cm]{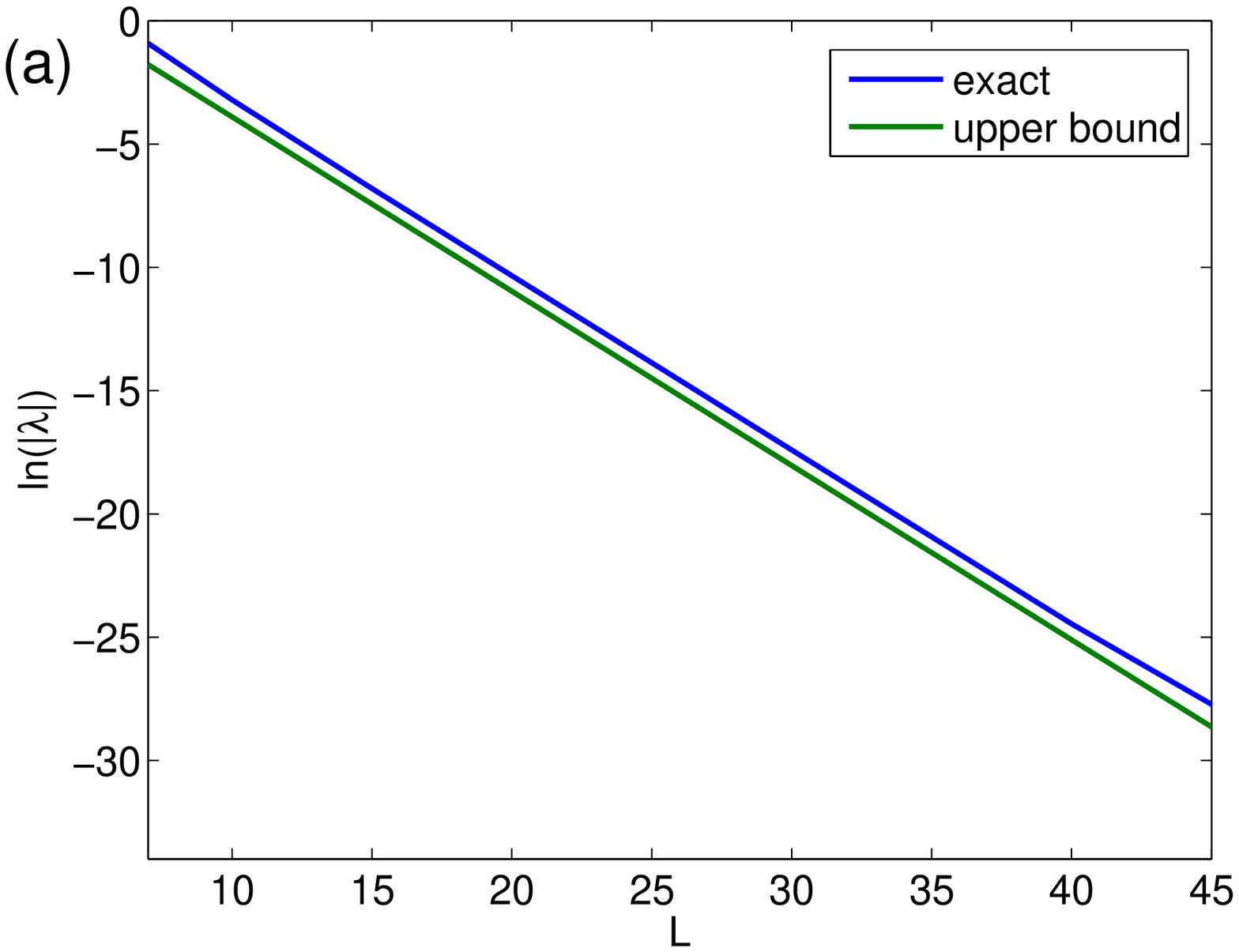}\includegraphics[width=5.5cm]{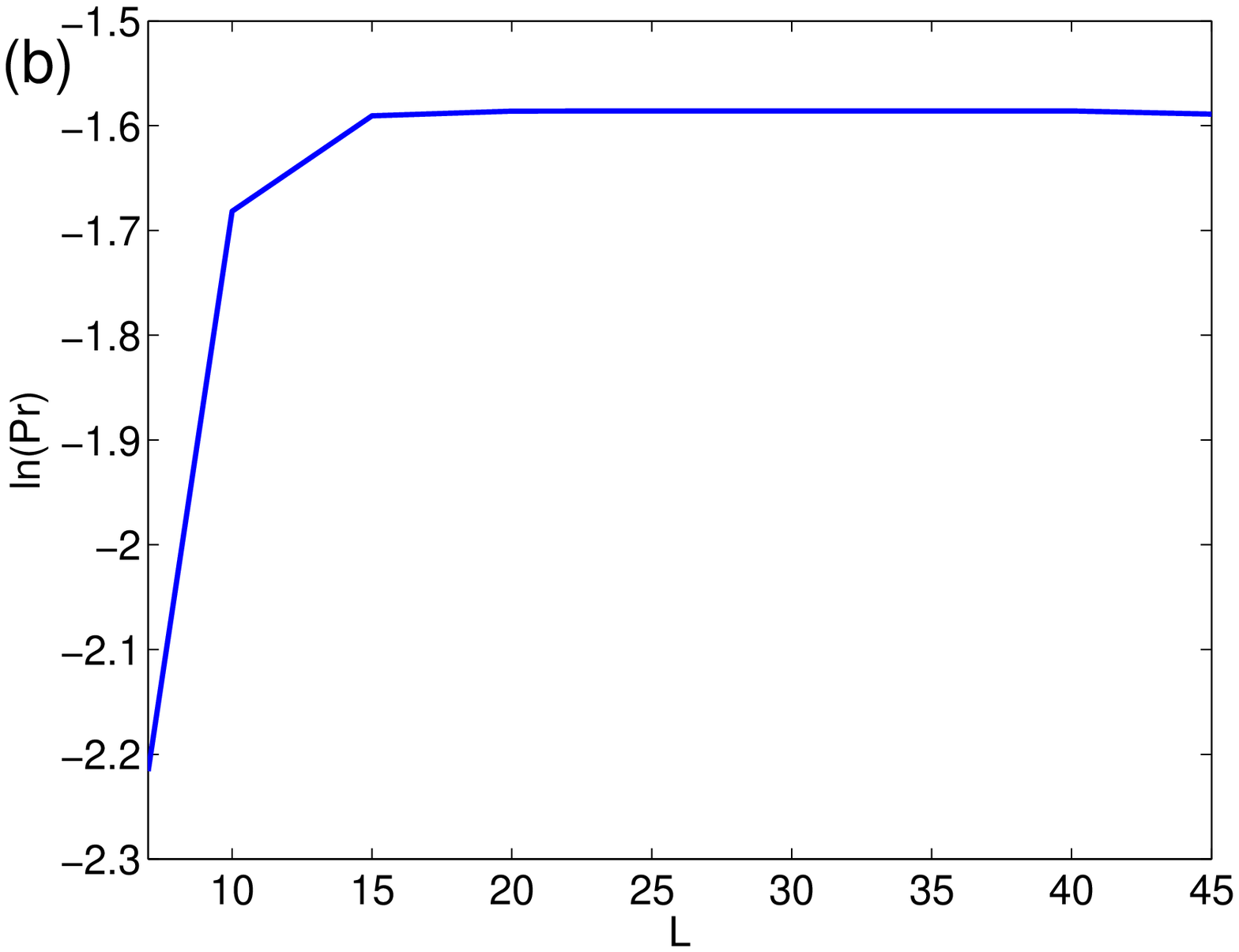}}
\caption{(a) : logarithm of the eigenvalue $\lambda_s$ and its approximation $\tilde{I}$ as a function of the size $L$. (b) : logarithm of the rest of the prefactor as a function of size.}\label{figapprox1}
\end{figure}

This eigenvalue $\lambda_s$ is the result of the eigenvalue problem
\begin{equation}
 \nabla^2 V|_{A_{s,0}}\Phi_i=-\partial_x^2 \Phi_i + (3A_{s,0}^2-1)\Phi_i=\lambda_{i,A_{s,0}} \Phi_{i,A_{s,0}}\,,
\label{eigvalprob_0}
\end{equation}
arising from the Hessian of $V$, with boundary conditions $\Phi_i(0)=\Phi_i(L)=0$. Since this operator is self adjoint, a good approximation is given using the min-max Rayleigh--Ritz theorem: the Rayleigh quotient gives an upper bound on $\lambda_s$. This upper bound can be as close as possible to $\lambda_s$. A analytical description can thus give a very good approximation of $\lambda_s$ and in particular of its decay rate. We detail the calculation in appendix~\ref{Ap_exponential_interactions}. At first order in $\exp(-L/\sqrt{2})$, the approximation reads
\begin{equation}
I\simeq-24e^{-\frac{L}{\sqrt{2}}}\,. \label{l1}
\end{equation}
The logarithm of the absolute value of $\lambda_s$ (calculated numerically) and $I$ are displayed in figure~\ref{figapprox1} (a). The ratio $I$ gives a very precise upper bound, the ratio between $\lambda_s$ and $I$ is approximately $2$, independently of $L$. This shows that the $\sqrt{2}$ factor in the exponential is exactly the decay rate of $\lambda_s$. Meanwhile, $24$ gives a reasonable order of magnitude of the factor in front of the exponential. Note that a scaling in $\exp(-L)$ had been calculated for the rate of evolution along the slow manifold of infinite size Allen--Cahn systems when two fronts are separated by a distance $L$ \cite{fh_fronts}.

Numerical computation of the whole spectrum of the Hessian of $V$ also allows us to compute the rest of the prefactor (Fig.~\ref{figapprox1} (b)). We can see that it quickly converges toward an asymptotic value independent of $L$. This confirms that the dependence on $L$ of the prefactor Eq.~(\ref{eqpref}) comes solely from $\lambda_s$. As a consequence, in the range of parameters in which the Eyring--Kramers law is valid, the mean first passage time will grow like $\exp(L/(2\sqrt{2}))$ with the domain size $L$.

\subsubsection{Duration of trajectories in one degree of freedom systems\label{sdur}}

The duration $s$ of a reactive trajectory is defined as the time elapsed between the instant where $A$ leaves $\mathcal{C}$ and the instant where $A$ touches $\partial \mathcal{B}$ (see Fig.~\ref{sketch}). For systems with only one degree of freedom, the distribution of durations of reactive trajectories
is known in the limit $\beta \to \infty$ (see \cite{asy}): it is a {\it Gumbel}
distribution with parameters $a$ and $b$
\begin{equation}\label{gumb}
\rho_{a,b}(s) = \frac{1}{a} {\rm exp}\left( -s_{ab} - {\rm exp}(-s_{ab})\right),
\tau \equiv {\rm mean} = b + \gamma a, {\rm var}^2 =  \frac{\pi^2 a^2}{6},
\end{equation}
with $s_{ab} = \frac{s-b}{a}$ and $\gamma \approx 0.6$ is the Euler constant.
The normalized distribution of the duration $\hat \rho (\hat s) \equiv
\rho( \frac{s-\tau}{{\rm var}})$ reads
\begin{equation}\label{ngumb}
\hat \rho(\hat s) = \frac{\pi}{\sqrt{6}} {\rm exp}\left( -\frac{\pi \hat s}{\sqrt{6}}-\gamma -
{\rm exp}(-\frac{\pi \hat s}{\sqrt{6}} -\gamma) \right).
\end{equation}
The average $\tau$ can be calculated as a function of $\beta$, $|\lambda_s|$ (the opposite of the second derivative of the potential at the saddle). In fact, it has been demonstrated by C\'erou \emph{et al.} that in the case of a one dimensional gradient system with a non degenerate saddle, we have in the $\beta\rightarrow \infty$ limit: $\tau=(\ln(\beta)+\ln(|\lambda_s|)+C_0)/|\lambda_s|$. In this result, the constant $C_0$ depends on the properties of the starting and arrival points \cite{asy}. In the one degree of freedom case, one can also show that the variance is  independent of $\beta$ and equal to $\pi/(\sqrt{6}|\lambda_s|)$.
Performing such an analysis in a many degrees of freedom system such as the Allen-Cahn equation is beyond the scope of this article, mainly because it requires a precise knowledge of the \emph{committor} function (see \S~\ref{Ad} as well as \cite{tpt,jcp,ons} and references within) in the neighbourhood of the saddle. This function of $A$ gives the probability that a realisation of the dynamics starting from a given $A$ reaches $\partial \mathcal{B}$ before $\partial \mathcal{A}$.

A transposition of the one degree of freedom result on the movement of the front on the saddle would lead us to expect that $\tau$ is affine in $\ln(\beta)$
\begin{equation}\tau\simeq\frac{\ln(\beta)+\ln(|\lambda_s|)+C_1}{|\lambda_s|}\,,\end{equation}
with $|\lambda_s|\propto e^{-L/\sqrt{2}}$, as approximated by $\tilde{I}$ Eq.~(\ref{l1}).

\subsection{Transition rates and reactive trajectories beyond Freidlin--Wentzell and Eyring--Kramers regimes}
\label{S3.3}

For large domain size $L$, fronts form in the bulk of the domain and have an effective interaction with the boundary and between themselves that decays exponentially with $L$. For small but finite noise, in the limit of large domains, we expect the front dynamics to be mainly diffusive.  In the infinite domain limit, it has actually been proven that the motion of the front is brownian with a diffusion coefficient $D=8/(3\beta)$ \cite{BBMP}.

We expect this diffusive front dynamics to be also relevant for reactive trajectories. This will be illustrated also by numerical computation in section \ref{S4}. This phenomenology contrasts with the classical Freidlin--Wentzell picture for which one expects the transition probability to be dominated by an action minimum contribution and small fluctuations around it.  Actually, as we explain in the following section, the saddle point approximation leading to a large-deviation result  is valid only in the parameter range $\beta\gg \exp(L/\sqrt{2})/L^2$. Similarly, the Eyring--Kramers formula for transition rates and mean first exit time is valid only in the same parameter range. Those remarks stress that the range of validity of the Freidlin--Wentzell regime and of the Eyring--Kramers formulas are quite limited for large domain sizes.

In this section we show that in the parameter range $\exp(L/\sqrt{2})/L^2 \gg \beta \gg 1/\Delta V$, even if the saddle point approximation fails, a large deviation result and an asymptotic formula for the transition rate can still be obtained through a specific approach. It is not a consequence of the Freidlin--Wentzell Principle of large deviations and the reactive trajectories are notably different from instantons. The transition rate differs from a Kramers--Eyring formula. We also discuss the typical duration of reactive trajectories and their distribution, which are markedly different from the classical ones in the Freidlin--Wentzell regime.

\subsubsection{Limits of Freidlin--Wentzell and Eyring--Kramers regimes\label{S3.3.1}}

In this section we discuss the range of parameters for which the Freidlin-Wentzell and Eyring--Kramers regimes are expected to be valid, first for a one dimensional stochastic differential equation, and then applied to the Allen--Cahn equation.

We consider the one dimensional dynamics $d\zeta/dt=-dV_e/d\zeta+\sqrt{2D}\eta(t)$. For this case, it is very classical to write quantities of interest in terms of explicit integrals. Those integrals can be evaluated asymptotically for small noise using a saddle point approximation \cite{gar}. For instance the Eyring--Kramers formula for the first passage time can be readily obtained this way. This derivation, or any other derivation, stress the importance of the dynamics close to the saddle point. The dynamics close to the saddle point can be described by  $d\zeta/dt=-\lambda_s \zeta+\sqrt{2D}\eta(t)$. We consider this equation with initial condition $\zeta(0)=0$. What is the average of the first hitting time $\tau$ for which $|\zeta|(\tau)=L/2$? What is a typical behavior of the trajectories leading to $|\zeta|(\tau)=L/2$ ; is it of a diffusive nature (the term $-\lambda_s$ is negligible), or is it rather dominated by the instability ($-\lambda_s$ dominates, except for very small $\zeta$)?  A simple dimensional analysis is enough to prove that the average of the first hitting time is $\mathbb{E}(\tau)=f(D/|\lambda_s| L^2)/|\lambda_s|$, where $f$ is a non dimensional function. In the small diffusion limit $D \ll |\lambda_s|L^2$, one recovers the behavior corresponding to the validity regime of the Eyring--Kramers relation. Then the distribution of $\tau$ is a Gumbel distribution, as discussed in section~\ref{sdur}, with $\mathbb{E}(\tau) = C_2\ln(D/|\lambda_s| L^2)/|\lambda_s|$, where $C_2$ is a non dimensional constant. In this case, after an initial short diffusive behavior, the trajectories follow an exponential escape. This regime corresponds also to the large deviation regime, where trajectories concentrate close to instanton--anti-instanton trajectories.  In the opposite limit $D \gg |\lambda_s|L^2$, the dynamics is dominated by diffusion. Then $\mathbb{E}(\tau) = C_3L^2/D$, with $C_3$ another nondimensional constant. In this regime the trajectory are diffusive and they do not have any concentration property.

In order to apply this discussion to the Allen--Cahn equation, we first note that in the case of large domains, the dynamics can be characterized by the motion of fronts. A single front motion is described by an effective diffusive dynamics in an effective potential $d\zeta/dt=-dV_e/d\zeta+\sqrt{2D}\eta(t)$, with $D=8/3\beta$. As discussed in section \ref{exponential_interactions}, the effective potential $V_e$ decays exponentially fast away from the boundary, then $\lambda_s = \lambda_0(L)\exp(-L/\sqrt{2})$. The factor $\lambda_0$ contains the exact value of the slowly varying factor of this eigenvalue (see \S~\ref{exponential_interactions}). From the previous discussion, we conclude that the large-deviation and Eyring--Kramers regime are observed for
\begin{equation}
\beta \gg \beta^\star(L)=\frac{\exp\left(\frac{L}{\sqrt{2}}\right)}{L^2|\lambda_0|}\,,\label{cond1}
\end{equation}
where $\sqrt{2}$ is the length scale  of the front, see for instance Eq.~(\ref{l1}). This is merely the explicit writing of the condition $D\ll |\lambda_s| L^2$. We can alternatively write this condition $L \ll L^\star(\beta)$, with $L^\star={\beta^\star}^{-1}(\beta)$ the unique root of this equation.

\subsubsection{Mean first passage times and transition rates for large domains durations of reactive trajectories}\label{S3.3.2}

We now consider large domains $L\gg1$ and the range of parameters $\exp(L/\sqrt{2})/L^2 \gg \beta \gg 1/\Delta V$ for which the Eyring--Kramers formula and the Freidlin--Wentzell phenomenology is not valid anymore. In that range of parameter, the dynamics is dominated by the nucleation of one or several fronts and their diffusive motion. In this subsection we discuss the simplest case, when a transition between the two attractors is due to the nucleation of a single front at one boundary, its diffusion throughout the domain until it reaches the other boundary to form the second attractor. The generalization of this discussion to the dynamics of multiple fronts would be rather easy. In section~\ref{numfront}, we will discuss the range of parameters for which the one front transitions are the typical ones. In this section, we propose a phenomenological theory of one front transitions, compute the transition rate (or mean first exit time) and discuss the statistics of durations of reactive trajectories.

We call  $\mathcal{A}$ and $\mathcal{B}$  the two attractors, and $P_{\mathcal{A}}$, resp. $P_{\mathcal{B}}$ the probability to observe them. We make several assumptions which are very natural. First we assume that when the system is in the attractor  $\mathcal{A}$ (resp.  $\mathcal{B}$) the nucleation of a front close to either one or the other boundary has a Poisson statistics characterized by a nucleation rate $\exp (-\beta\Delta V)/\chi$, with $\chi$ a time scale that has a finite limit in the limit of small $\beta\Delta V$. This is natural as the nucleation is an activation process. Moreover it is natural to assume that $\chi$ has a finite limit for $L$ going to infinity. Secondly, we assume that in the asymptotic large $L$ limit,  the dynamics of the front is the same as in an infinite domain, studied for instance by Brassesco \emph{et al.} \cite{BBMP}. The front then diffuses like a Brownian particle with diffusion coefficient $D=8/(3\beta)$. Let us call $P(x,t)$ the density of the probability to observe a front at distance $x$ (resp. $x-L$) from its nucleation boundary if it has been nucleated from the attractor $\mathcal{A}$ (resp. $\mathcal{B}$). We note that a front can be nucleated either at the right or the left boundary, from either the attractor $\mathcal{A}$ or the attractor $\mathcal{B}$. However, because of the symmetry of the problem, and because we assume that their is a single front at a time, the probability $P(x,t)$ defined that way is enough to characterize the state of the system.

A key point is to relate the boundary condition for $P(x,t)$ with $P_\mathcal{A}$ and $P_\mathcal{B}$. The process at the boundary involves the creation and the destruction of a front. In order to model these phenomena, we denote $v$ the rate per unit of length for a front in the vicinity of the position $x=0$  disappears and form the attractor $\mathcal{A}$. By symmetry, this velocity $v$ is also the rate of disappearance per unit of length of a front close to $x=L$ to form attractor  $\mathcal{B}$. It is natural to assume that $v$ has a finite limit when $L$ goes to infinity, as the effect of the second boundary becomes negligible. The phenomenological equations then read
\begin{align}
\frac{dP_{\mathcal{A}}}{dt}=-\frac{\exp (-\beta\Delta V)}{\chi}P_{\mathcal{A}}+vP(0,t) \,, \label{PA}  \\ \frac{dP_{\mathcal{B}}}{dt}=-\frac{\exp (-\beta\Delta V)}{\chi}P_{\mathcal{B}}+vP(L,t)\,, \label{PB}
\end{align}
coupled to the Fokker--Plank equation
\begin{equation}
\frac{\partial P}{\partial t}=D\frac{\partial^2P}{\partial x^2}, \label{PKplateau}
\end{equation}
with boundary conditions at $x=0$ and $x=L$
\begin{align}
D\frac{\partial P}{\partial x}(0,t)=-\frac{\exp (-\beta\Delta V)}{\chi}P_{\mathcal{A}}+vP(0,t) \ {\rm and} \label{BP1} \\
D\frac{\partial P}{\partial x}(L,t)=\frac{\exp (-\beta\Delta V)}{\chi}P_{\mathcal{B}}-vP(L,t) \label{BP2} \,.
\end{align}

The equilibrium distribution should be the one corresponding to the Gibbs distribution for the Allen-Cahn model. In the limit of infinite domain sizes, we expect, at equilibrium $P_{\mathcal{A}}=P_{\mathcal{B}}\sim1/2$, and $P(x)=\exp(-\beta\Delta V)/l_1$, where $l_1$ is a length of order of the front size that could be computed easily from the Gibbs distribution using a saddle point approximation. Looking at the equilibrium of the phenomenological model (\ref{PA}-\ref{BP2}), we conclude that the relation $l_1=2v\chi$ must hold. We note that our phenomenological model verifies detailed balance with respect to its invariant measure.

While the value of $D$ has been computed from the Allen-Cahn equation, the values of $v$ and $\chi$   remains phenomenological parameters so far. It would be a very interesting problem to compute them from the Allen-Cahn equation (as explained above they are related through  $l_1=2v\chi$).\\

We first compute the flux $j\equiv -D \partial P/\partial x$  for a steady solution of this phenomenological model for which we impose $P_{\mathcal{A}}=1$ and $P_\mathcal{B}=0$ (we consider equations~(\ref{PKplateau},\ref{BP1},\ref{BP2}) with $P_{\mathcal{A}}=1$ and $P_\mathcal{B}=0$). From the steadiness of the process and Eq.~(\ref{PKplateau}), we have that $P(x)=-jx/D+P(0)$. The flux $j$ and $P(0)$ are determined from the boundary conditions Eq.~(\ref{BP1}) and Eq.~(\ref{BP2}).  We obtain
\begin{align*}
j = \frac{D}{\omega(vL+2D)}\exp(-\beta \Delta V) \underset{L\gg1}{\sim} \frac{16}{\beta l_1L}\exp(-\beta \Delta V),
\end{align*}
where we have used for the left hand side $D=8/3\beta$ and $2v\chi= l_1$.

Such a steady solution is established after a diffusion characteristic time $\tau_D=L^2/D=3\beta L^2/8$. Whenever this diffusion time is much smaller  than the nucleation time $\chi_{{\rm n}}=\chi\exp (\beta\Delta V)$, it is easily understood that the evolution of $P_{\mathcal{A}}$ and $P_{\mathcal{B}}$ is described by an effective slow equation. This slow equation can be easily derived. Hence for $1 \ll L \ll \sqrt{\chi/\beta}\exp(\beta \Delta V/2)$, we obtain
\begin{align}
\frac{dP_{\mathcal{A}}}{dt}=-\mu(P_{\mathcal{A}}-P_{\mathcal{B}})\,{\rm and}\\ \frac{dP_{\mathcal{B}}}{dt}=-\mu(P_{\mathcal{B}}-P_{\mathcal{A}})\,,
\end{align}
where $\mu$ takes the asymptotic value of the current $j$ computed above
\begin{align}
\mu = \frac{16}{\beta l_1L}\exp(-\beta \Delta V)\,. \label{lambda}
\end{align}
As a conclusion, whenever $1 \ll L \ll \sqrt{\chi/\beta}\exp(\beta \Delta V/2)$,  the dynamics of the transition from the attractor $\mathcal{A}$  to the attractor  $\mathcal{B}$ is an effective Poisson process, with transition rate $\mu$. The mean first exit time from the basin of attraction of one of the attractors  is then $1/\mu$.

We do not discuss the parameter range $L \gg \sqrt{\chi/\beta}\exp(\beta \Delta V)$, because the hypothesis of a one front solution is no more valid in that case. This limit is studied in section~\ref{numfront}. We discuss in next section the range of parameters for which the one front states are actually the typical ones for reactive trajectories.

We have discussed in this section a phenomenological model of the kinetics of a one front reactive trajectory. However, it should be noted that this model can be easily extended to a kinetic model including the possibility to have $n$ fronts at the same time.

\subsubsection{Duration of reactive trajectories for large domains}\label{S3.3.3}

We now discuss the duration of reactive trajectories and come back to the range of validity of our hypothesis that only one front forms, always for large domain sizes.

As it is clear from the phenomenological model described in the previous section, the kinetics of the transitions are governed by diffusion in the large size limit.  For random walk in dimension 1, it can be demonstrated that the average duration of reactive trajectories is proportional to the square of the size of the domain and inversely proportional to the diffusion coefficient \cite{asy}. We then have
\begin{align}
\tau = C_4\beta L^2\label{taudiff}\,,
\end{align}
where $c$ is a non-dimensional constant. Note that in the case of a free random walk with one degree of freedom, the expected pdf of trajectory durations is not a Gumbel distribution any more. Moreover, the variance depends on $\beta$. In fact, it is proportional to the average duration ${\rm var}= \tau\sqrt{2/5}$ \cite{asy}.

The one front case will remain the typical transition state of the system only if the average duration of reactive trajectories $\tau = C_4\beta L^2$ is much smaller than the nucleation time $\chi_{\rm n}=\chi\exp (\beta\Delta V)$.  We conclude that this is true whenever
\begin{align*}
L \ll \sqrt{\chi/\beta}\exp\left(\frac{\beta \Delta V}{2}\right)\,.
\end{align*}
This criterion turns out to be the same as the one derived for the validity of effective Poisson process for describing transitions from one state to another. The same criteria will be also obtained in the next section using a purely equilibrium argument (from the Gibbs distribution).

The length $L$ present in the transition rate $\mu$ (Eq.~\ref{lambda}) or in the duration $\tau$ is that of the potential plateau. It is the length of the domain minus the size of the two boundary layers on both ends $L-2\delta L$. An analytical estimate of $\delta L$ can be performed by calculating $1/(\partial V/\partial y)_{y=0}$, where $y$ is the position of the front, using the ansatz of equation~\ref{ansatzfront} (see \S~\ref{Ap_exponential_interactions}) and letting $L\rightarrow \infty$. The calculation is very similar to that of the potential difference $\Delta V$. This gives a value of $8$, while a more precise numerical calculation shows that $\delta L\simeq 5$.

\subsubsection{Number of fronts \label{numfront}}
At finite-temperature, the system may have several fronts separating the domain into different parts. In order to calculate how many,
a useful analogy is that of a gas of weakly interacting subdomain fronts \cite{wall_pre}.
We can consider the probability $p_n$ of having $n$ fronts in the domain relatively
to the probability $p_0$ to be in one of the minima $A_0^\pm$. We assume that the
fronts do not interact at all. This assumption is realistic since in practice the interaction
amplitude decreases exponentially with the distance between the fronts
\cite{bray,wall_pre}. The probability of having $n$ fronts relative to $p_0$ reads
\begin{equation}
p_n = 2p_0 \int_{X_1=0}^L dX_1 \int_{X_2=X_1}^L dX_2 \cdots \int_{X_n = X_{n-1}}^L dX_n
{\rm exp}(-\beta \Delta V_n).
\end{equation}
Here the factor 2 accounts for the natural symmetry $A \to -A$ and the integration is over all
possible positions $X_1,\cdots,X_n$ of the fronts, the term $\Delta V_n = n \Delta V_1$
(see Eq.\ref{potdiff}) and is constant, i.e. independent of the front positions. The calculation
of the above integral gives at the end
\begin{equation}
p_n = 2p_0  \frac{L^n}{n!} {\rm exp}(-n\beta \Delta V_1).
\end{equation}
Note that the factorial term $1/n!$ differs from earlier predictions \cite{otto}.
It means that fronts behave in fact like identical particles in this treatment
and it is often found in classical statistical physics.
We now identify the regions of the plane $(L,\beta)$ for which observing $n$ random-walk fronts
is the more likely. It is simply done by identifying the curves $p_n = p_{n-1}$ (when having $n$ fronts becomes more probable than $n-1$) and
$p_n = p_{n+1}$ (when having $n+1$ fronts becomes more probable than $n$). We find
\begin{equation}\label{cond2}
L {\rm exp}(-\beta \Delta V_1) - 1\leq n \leq L {\rm exp}(-\beta \Delta V_1).
\end{equation}
It gives the size of the coherent domain $L/n$ which is fixed by the competition
between the level of noise present in the system and the potential cost of creating fronts.

\subsection{Theoretical phase diagram for reactive trajectories}
\label{S3.4}
\begin{figure}
\centerline{
\includegraphics[width=10cm]{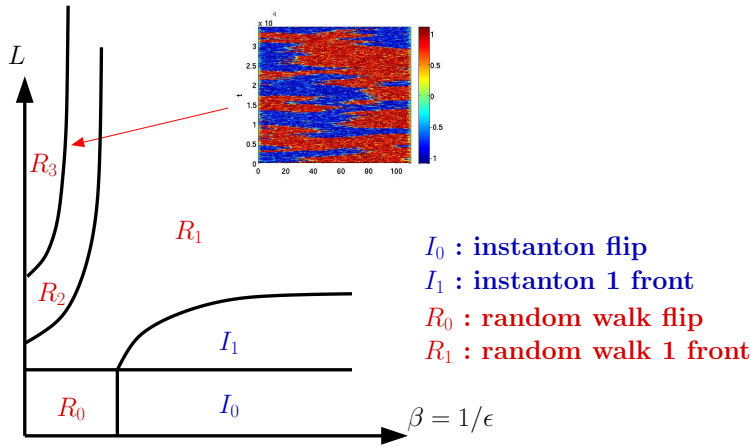}
}
\caption{Sketch of the phase diagram of possible reactive trajectories in the $(\beta,L)$ plane.
A random walk solution with one or two front is shown for illustration.}
\label{diag1}
\end{figure}

We provide here a compilation of the results we found previously. Figure \ref{diag1}
shows the type of reactive trajectory as a function of the inverse temperature $\beta$ and
domain size $L$. The different curves in this parameter space are found from
conditions (\ref{cond1}),(\ref{cond2}) as well as the results of Fig.\ref{deltav}.
Note that condition (\ref{cond1}) being exponential in $\beta$, the existence of genuine
instantons is concentrated in a very narrow band of relatively small size $L$, unless $\beta$ is
very large. Similarly, condition (\ref{cond2}) indicates that reactive trajectories with
several fronts exist in a very narrow band of inverse temperature $\beta$.
Note also that in practice, most of the parameter plane is occupied by random walk type
of trajectories.

\section{Numerical results}
\label{S4}
We now present systematic computations of reactive trajectories using AMS algorithm.
We discuss first the qualitative properties of these trajectories in \S~\ref{S4.1}.
We then discuss in a more quantitative way the trajectory properties and compare them with
the asymptotic theoretical predictions. We decompose this discussion into three parts:
mean first passage time (\S~\ref{S4.2}), distribution of the duration of reactive trajectories (\S~\ref{S4.4}) and escape probability (\S~\ref{S4.3}). The numerical procedure is described in the appendices~\ref{Ad} and~\ref{Ac}. Along with the details on the algorithm, we explain the importance of the number of trajectories $N$ and the number of realisations of the procedure on convergence in these sections.
\subsection{Reactive trajectories}\label{S4.1}
We consider here four different domain sizes $L$. We use $N=1000$ trajectories in these simulations: this number is small enough for the simulations to be quick, but large enough for the outputs to be reliable.

\begin{figure}
\centerline{
\includegraphics[width=18cm]{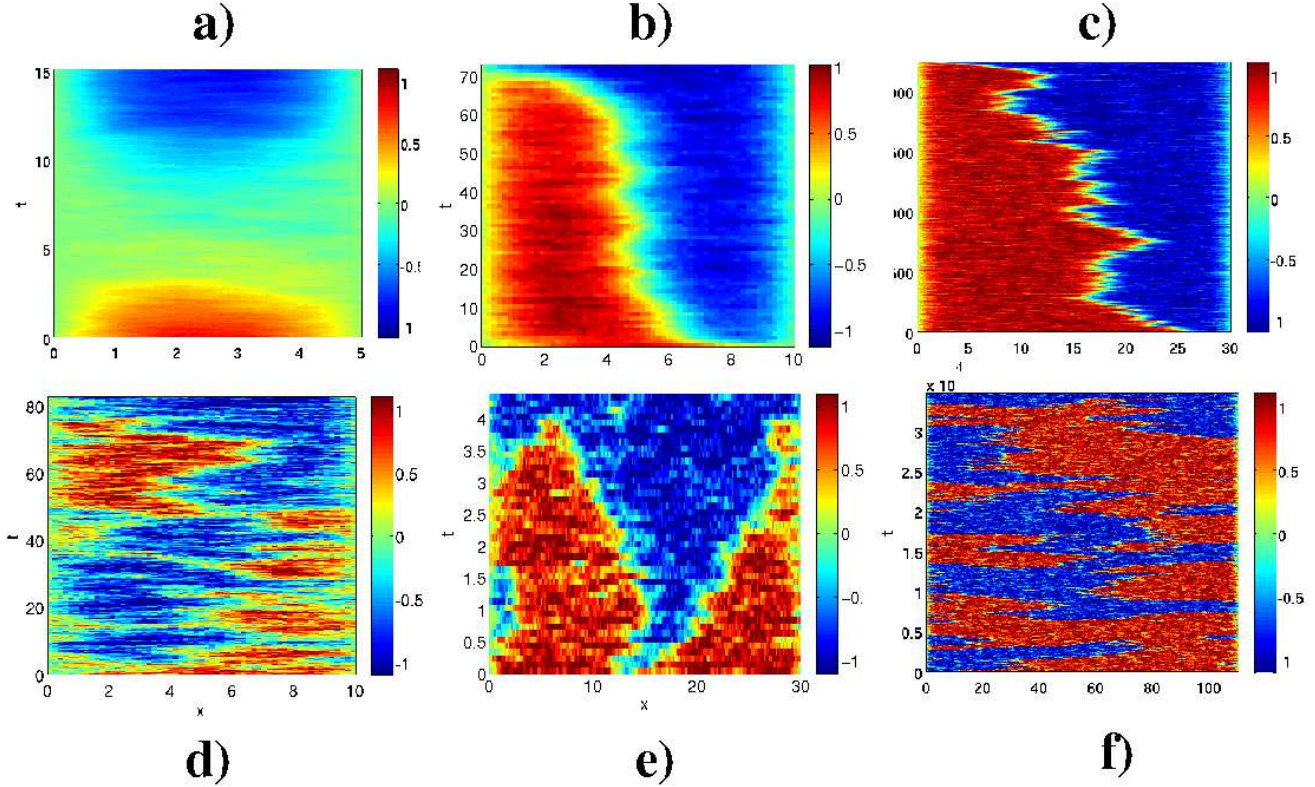}
}
\caption{Examples of reactive trajectories computed using AMS algorithm for
$N=1000$, $dt=10^{-3}$ and $dx=1/18$. All the trajectories are typical of the average duration.
a): $L=5$, $\beta=300$. b): $L=10$, $\beta=150$. c): $L=30$, $\beta=30$. d): $L=10$, $\beta=5$.
e): $\beta=7$, $L=30$. f): $\beta=7$, $L=110$.}
\label{possol}
\end{figure}

For $L=5$ and $\beta=300$,
the transition rate $1/T$ is estimated from numerical results (See Appendix \S~\ref{sout}, Eq.~\ref{mfpt_ams}). In that case, it is $1/T \simeq 3.7 \cdot 10^{-49}$. Note that in this system, the relaxation time is of order $1$: this is much smaller than the mean first passage time that we computed. A typical reactive trajectory is
shown in Fig. \ref{possol}a. It corresponds to the first type of (flip) instanton going through
the saddle $A_0 = 0$ due to the relatively small size $L$.
It is qualitatively comparable with the flip instanton shown in Fig. \ref{figinst}a).

Another regime is for larger $L$. For $L=10$ and $\beta=150$ the transition rate
is $1/T \simeq 3.6 \cdot 10^{-67}$. In this case, one finds reactive trajectories with only
one front (see Fig. \ref{possol}b). It is consistent with the theoretical results where
$A_1$ becomes the saddle with lowest potential (see \S~\ref{S3.2} and Fig. \ref{deltav}b).
The front propagates quickly from $x=0$ to $x=L/2$ spends a very long time there and then
moves forward to $x=L$. The main difference with a genuine front instanton (Fig. \ref{figinst}b)
is that the position of the front fluctuates slightly around $x=L/2$. For these order of magnitude of the rate of escape, it is clear that a numerical approach such as AMS is necessary. Indeed, for spacially extended systems, direct numerical simulations are conceivable up to durations $t\lesssim 10^8$. Computing mean first passage time that are much larger than that is not feasible.

We then consider a domain of size $L=30$ with $\beta=30$ which yields
$1/T \simeq 8.3 \cdot 10^{-16}$.
The algorithm finds reactive trajectories with one front as shown in Fig. \ref{possol} (c).
At this rather low $\beta$, the solutions present some significative differences with
front instantons (Fig. \ref{figinst}b). Although the transients from $x=L$ to
$x \sim L_1 = 25$ and that from $x\sim L_2 = 5 $ to $x=0$ are very short as expected from
deterministic relaxations, the front propagation has large fluctuations in the interval $[L_2,L_1]$.
This is reminiscent of the three-stage process described in section \ref{S3.3.2}. Note that this provides us an example of the potential boundary layer thickness $\delta L\simeq 5$ introduced in section~\ref{S3.3.3}.
At this low $\beta$ regime, saddle-point approximations of Freidlin--Wentzell and Eyring--Kramers
theories are no more valid. In fact, most of the reactive trajectories found by the algorithm
are not instantons when $\beta$ becomes small. A typical case is shown in Fig.\ref{possol}d
for $\beta=5$ and $L=10$ where $1/T \simeq 3 \cdot 10^{-3}$. One still observes one front
solution but with higly random motion. Another case at $\beta=7$ and $L=30$
($1/T \simeq 4.3 \cdot 10^{-4}$),
exhibits two fronts with rather regular propagation (see Fig.\ref{possol}e). This is
comparable with trajectories minimizing the action at small duration $T$ constraint \cite{VE}.
Note that even at low $\beta$, these two fronts trajectories are only a minor part of the
fastest trajectories.
Finally, we consider the case of a very large domain with $L=110$ and $\beta = 7$
($1/T \simeq 1.3 \cdot 10^{-5}$). In this case, one observes several isolated fronts with random
walk propagation until the system reaches $A_0^-$. Note that this situation only occurs for
$\beta \le 7$ and $L\gtrsim 30$. These values are critical in the sense that
for $\beta > 7$ and $L < 30$, reactive trajectories with a single front reappear.

These first simulations show us that instantons are observed for large $\beta$. The other expected diffusive regimes also found. Most of the relevant sets of parameters lead to transition rate so small that they cannot be attained using direct numerical simulations: methods such as adaptive multilevel splitting are of great help in such studies.

\subsection{Mean first passage time}\label{S4.2}

\begin{figure}
\centerline{\includegraphics[height=4cm]{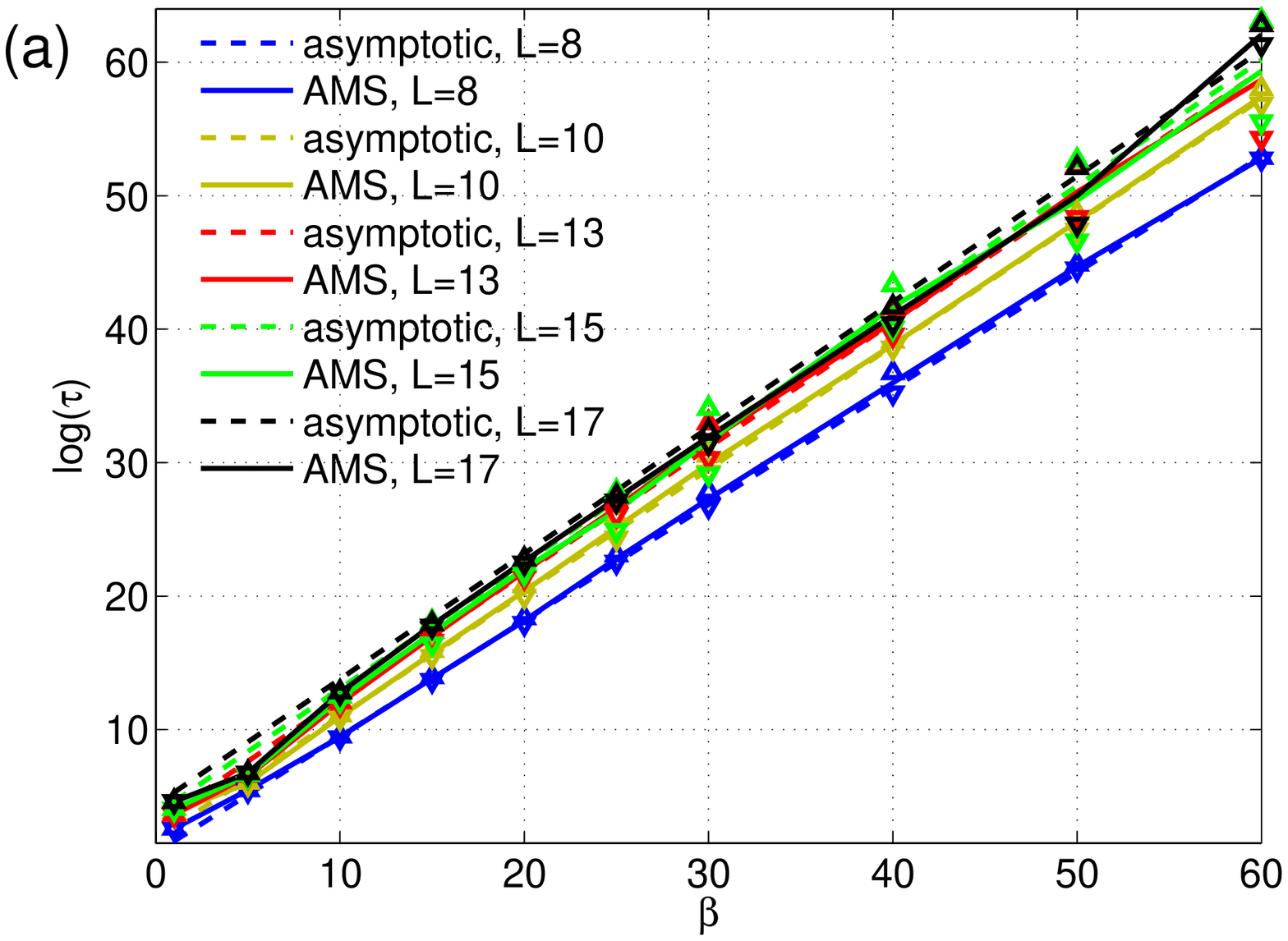}\includegraphics[height=4cm]{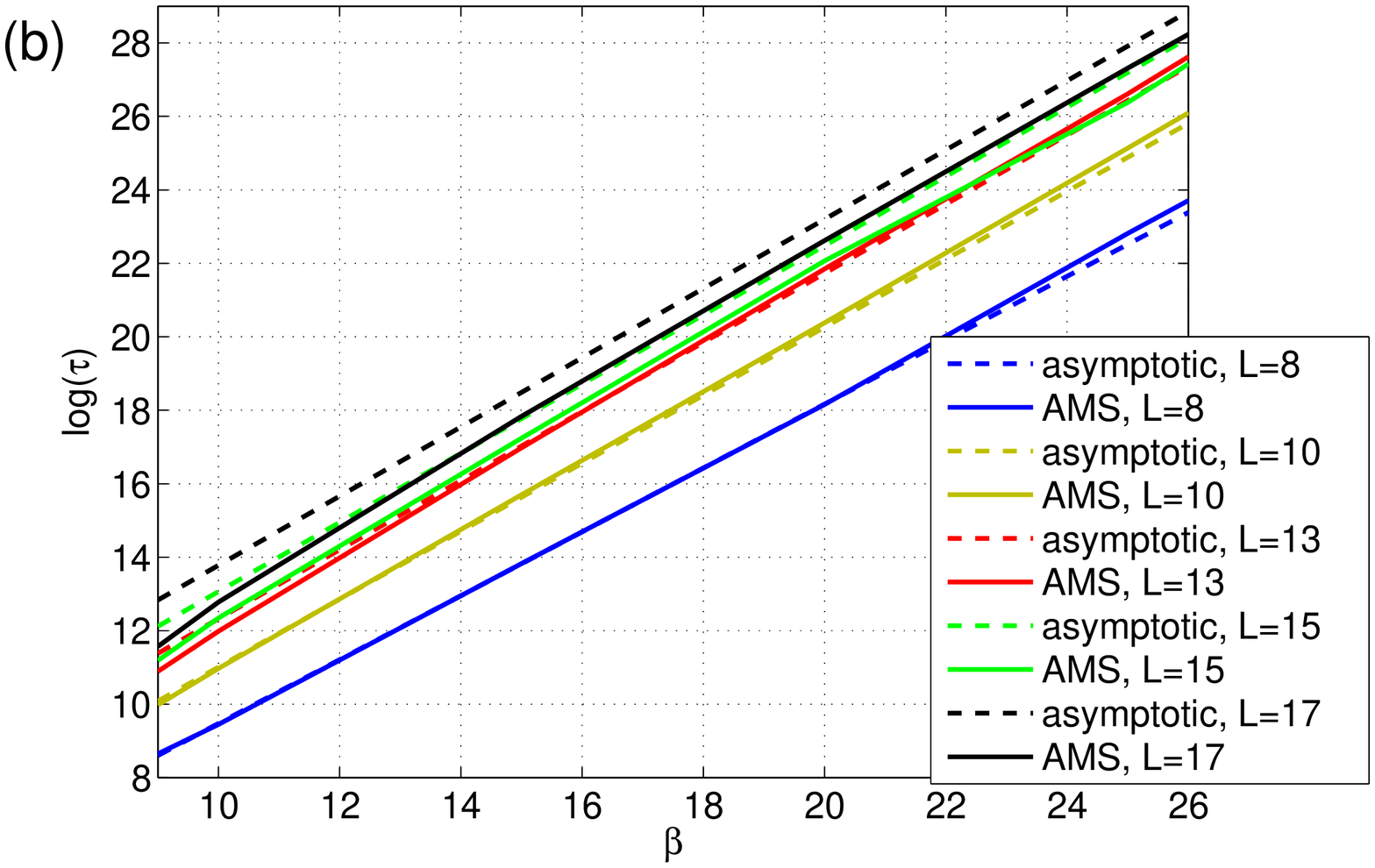}\includegraphics[height=4cm]{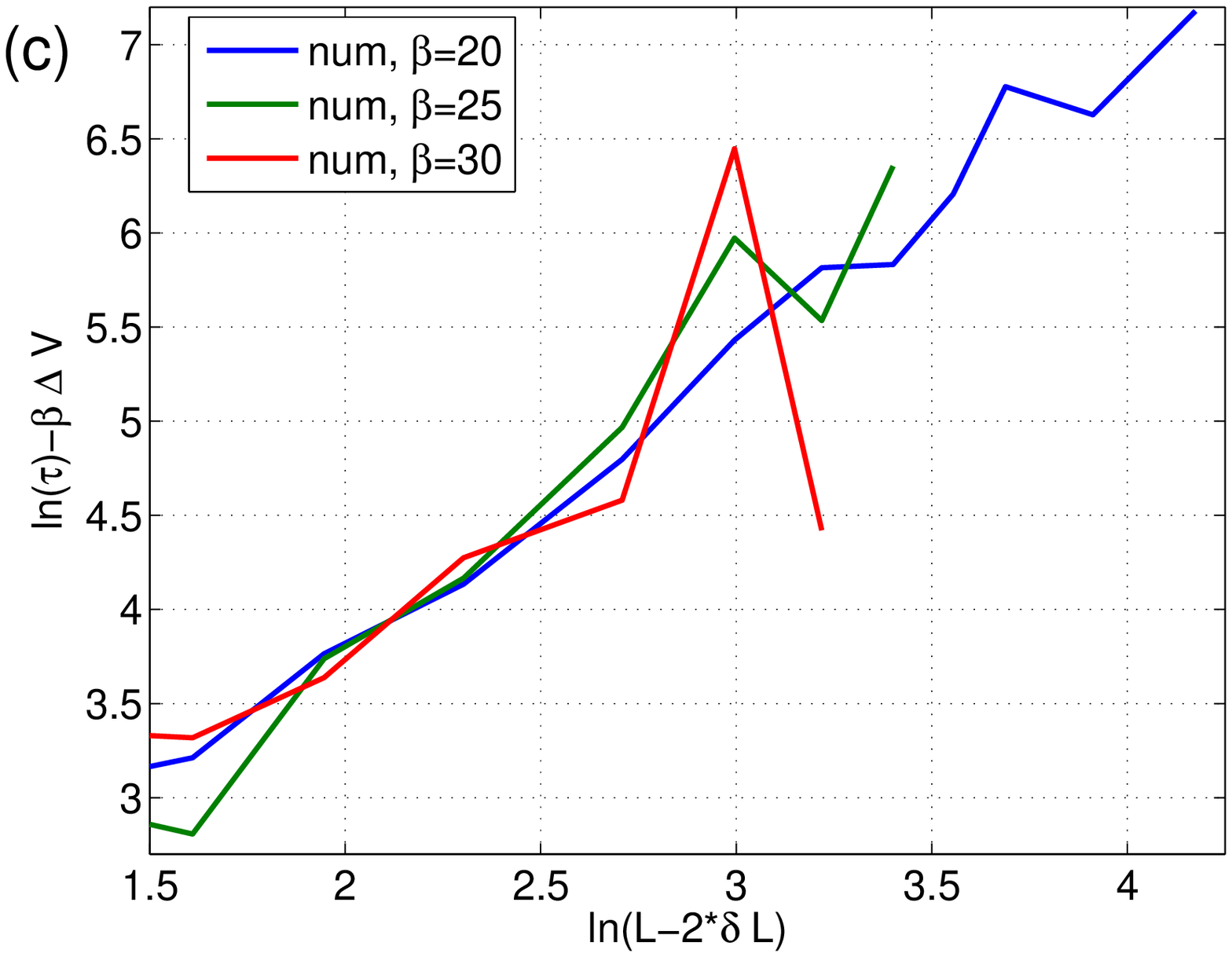}}
\caption{(a): Logarithm of the mean first passage time computed using
the asymptotic formula (Eq.\ref{mfpt}) and using the AMS one (Eq.\ref{mfpt_ams})
as a function of $\beta$. (b): Logarithm of the mean first passage time as a function of $\beta$ zoomed in the interval $9\le \beta \le 26$. (c): Logarithm of the mean first passage time (analytical and numerical result) minus $\Delta V\beta$ as a function of $\ln(L-2\delta L)$ for several values of $20\le\beta\le 30$.}
\label{mfpt_fig}
\end{figure}

An efficient way to test the validity of our results is to compare quantities which
can be precisely predicted. In particular, the mean first passage time gives us a good
benchmark (see Eq. (\ref{mfpt}) and the algorithmic formula used in Eq. (\ref{mfpt_ams}), \S~\ref{sout}).
We thus compute $T$ as a function of $\beta$ and $L$ for $\beta \in [1,60]$ and $L \in [8,17]$.
In this regime, the saddle is $A_1$. We perform 10 independent realisations of
the algorithm for $N=1000$ and
$N=10~000$ in the case $L=17$ and perform an average over these realisations.
The results are shown in figure~\ref{mfpt_fig} (a). We superimpose the corresponding analytical predictions described in section~\ref{S3.1}. In practice, the numerical results given by AMS are estimates of the mean first passage time,
trajectory duration, \emph{etc}. As detailed in appendices~\ref{Ad} and~\ref{Ac}, the actual values lie in an interval of confidence around this estimate. Using the mathematical convergence results \cite{cg07,STCO,poi} and propagation of errors, we construct such intervals of confidence for the mean first passage times we computed and add them in the form of error bars for figure~\ref{mfpt_fig} (a). We took care to use enough trajectories $N$ and enough realisations of the algorithm for these interval of confidence to be small, explaining why error bars are not visible in the figure. In the rest of the article, these intervals of confidence will not be included for readability.

We move to the discussion of the physical content of the numerical results. We can see in figure~\ref{mfpt_fig} (a) and more particularly in the zoom of figure~\ref{mfpt_fig} (b) that for $\beta >7$ and $L\le 13$, we have a near exact match between the mean first passage time estimated by AMS and the analytical result of the Eyring--Kramers formula. The ratio of the numerical estimate over the analytical prediction is $1\pm 0.1$. This means that in this part of the parameter plane, the transition rate $1/T$ decays exponentially with $\beta$ at a rate $-\Delta V$ and also exponentially with $L$ at a rate $-1/\sqrt{2}$. The Eyring--Kramers formula is an asymptotic result in the large $\beta$ limit, so that the small discrepancy at small $\beta$ is not surprising (this will be more visible when we examine the results on the escape probability in section~\ref{S4.3}). In section~\ref{S3.3} we explained why the Eyring--Kramers is not valid for larger sizes. We expected that the agreement between numerical estimations of the mean first passage time and the prediction of the Eyring--Kramers formula would only be valid up to a certain size. The value $L=13$ gives us a lower bound on the length $L^\star(\beta)$ above which these large $\beta$ result should not be expected. This lower bound is in practice independent on $\beta$: this is in perfect agreement with the prediction that $L^\star$ grows logarithmically with $\beta$ (see Eq.~(\ref{cond1})).

For $L\ge 15$, the numerical estimate of the mean first passage time is smaller than the value given by the Eyring--Kramers formula as shown in figure~\ref{mfpt_fig} (b). This indicates that $L$ has
crossed $L^\star$ for this range of $\beta$. Nevertheless, we find that the Freidlin--Wentzell
large-deviation prediction is very robust although the Eyring--Kramer prefactor is
no more valid \cite{ht}. The logarithm of the mean first passage time still grows linearly with $\beta$, with the same slope $\Delta V$ as the asymptotic Eyring--Kramers result. Meanwhile the ratio of the numerical estimate over the analytical result is approximately $0.5$ for $L=15$ and approximately $0.4$ for $L=17$. This value is independent on $\beta$. This numerical result is reminiscent of the crossover between an exponential growth of $T$ with $L$ (see Eq.~(\ref{lambda}) \S~\ref{exponential_interactions}) and a much slower polynomial growth of $T$ with $L$ (see Eq.~(\ref{l1}), \S~\ref{S3.3.2})

We examine the large size $L$ regime in more detail in view of the analytical results of section~\ref{S3.3.2} (Eq.~(\ref{l1})). In the diffusive regime, this formula tells us that $\ln(T)-\beta \Delta V=\ln(L-2\delta L)+\ln(\beta l_1/16)$. The boundary length $\delta L$ had been introduced in section~\ref{S3.3.3} to account for the actual length of the potential plateau. We display $E\equiv \ln(T)-\beta \Delta V$ as a function of $\ln(L-2\delta L)$ in figure~\ref{mfpt_fig} (c), where $T$ is estimated using the AMS and $\Delta V$ is computed numerically. Three values of $\beta=20$, $25$ and $30$ were considered. We can first see that $E$ is independent on $\beta$, confirming that $T$ grows exponentially with $\beta$, whatever the size. We  find an affine growth of $E$ with $ln(L-2\delta L)$. This indicates that in the large size $L$ regime, the mean first passage time $T$ grows as a power of $L$. This finding differs quite strongly from the exponential growth expected from the Eyring--Kramers formula. A fit of the numerical data gives an estimate of $1.5$ for the slope of $E$ as a function of $\ln(L-2\delta L)$.

\subsection{Duration of reactive trajectories}\label{S4.4}

\begin{figure}
\centerline{\includegraphics[width=6cm,clip]{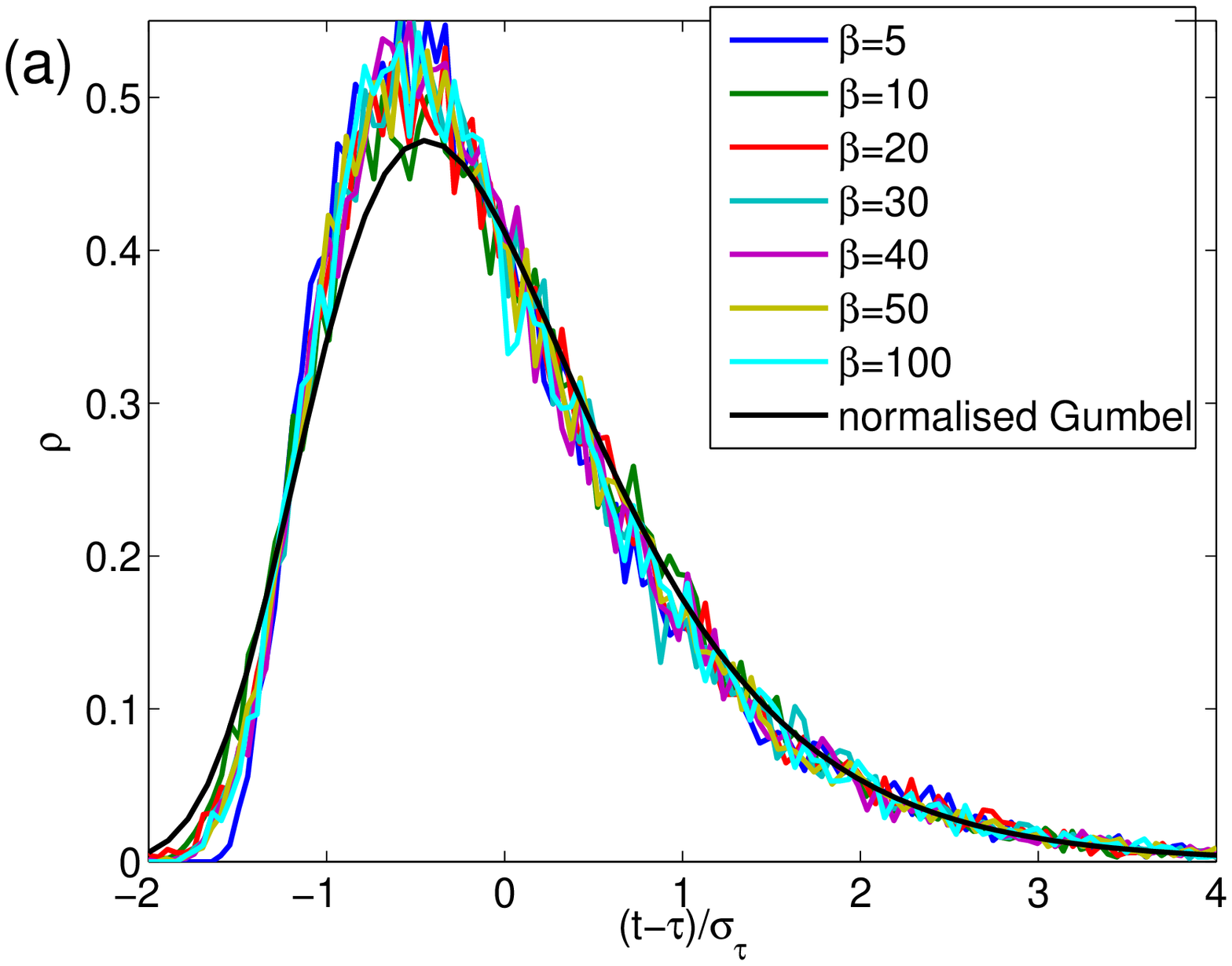}\includegraphics[width=6cm]{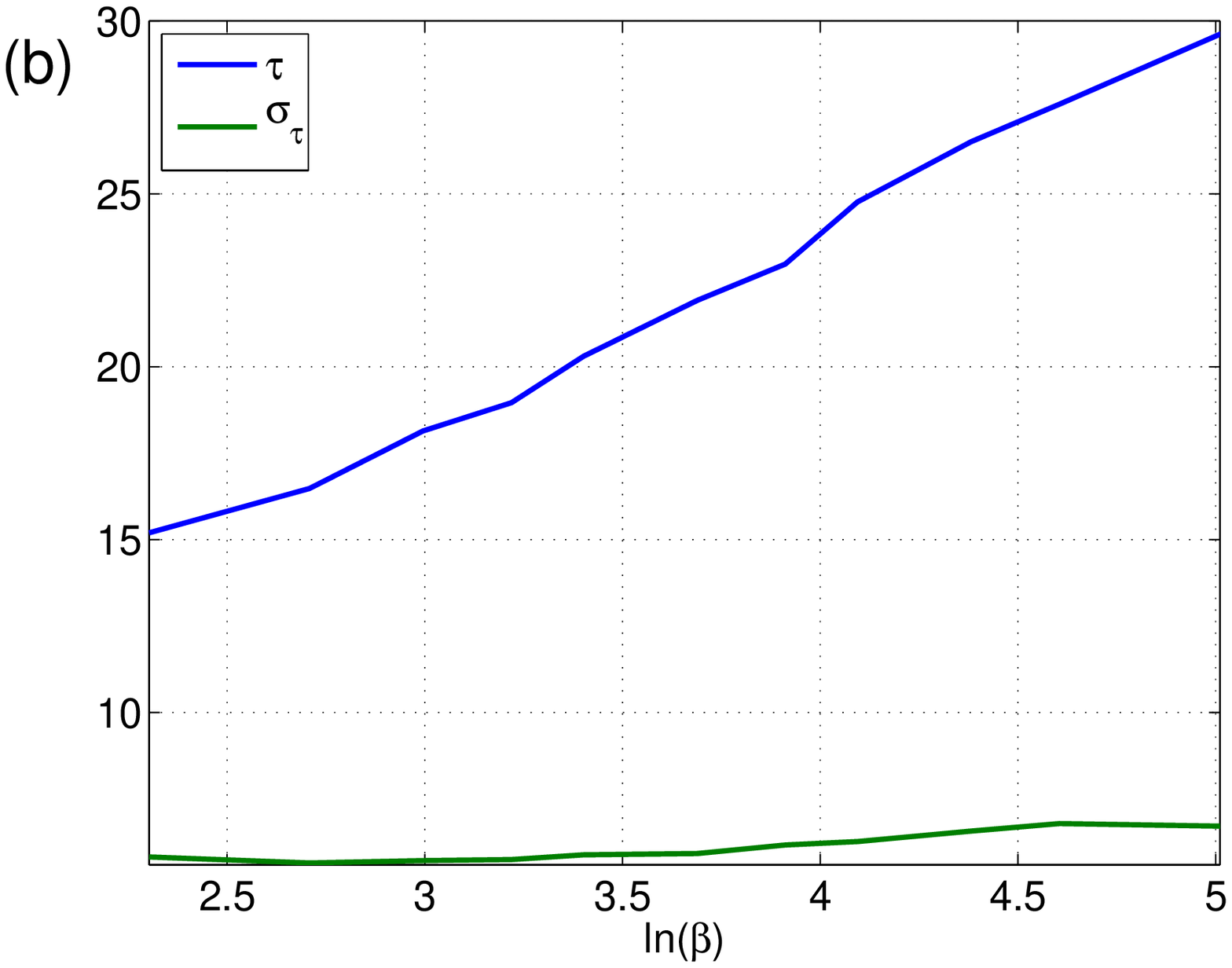}}
\centerline{\includegraphics[width=6cm,clip]{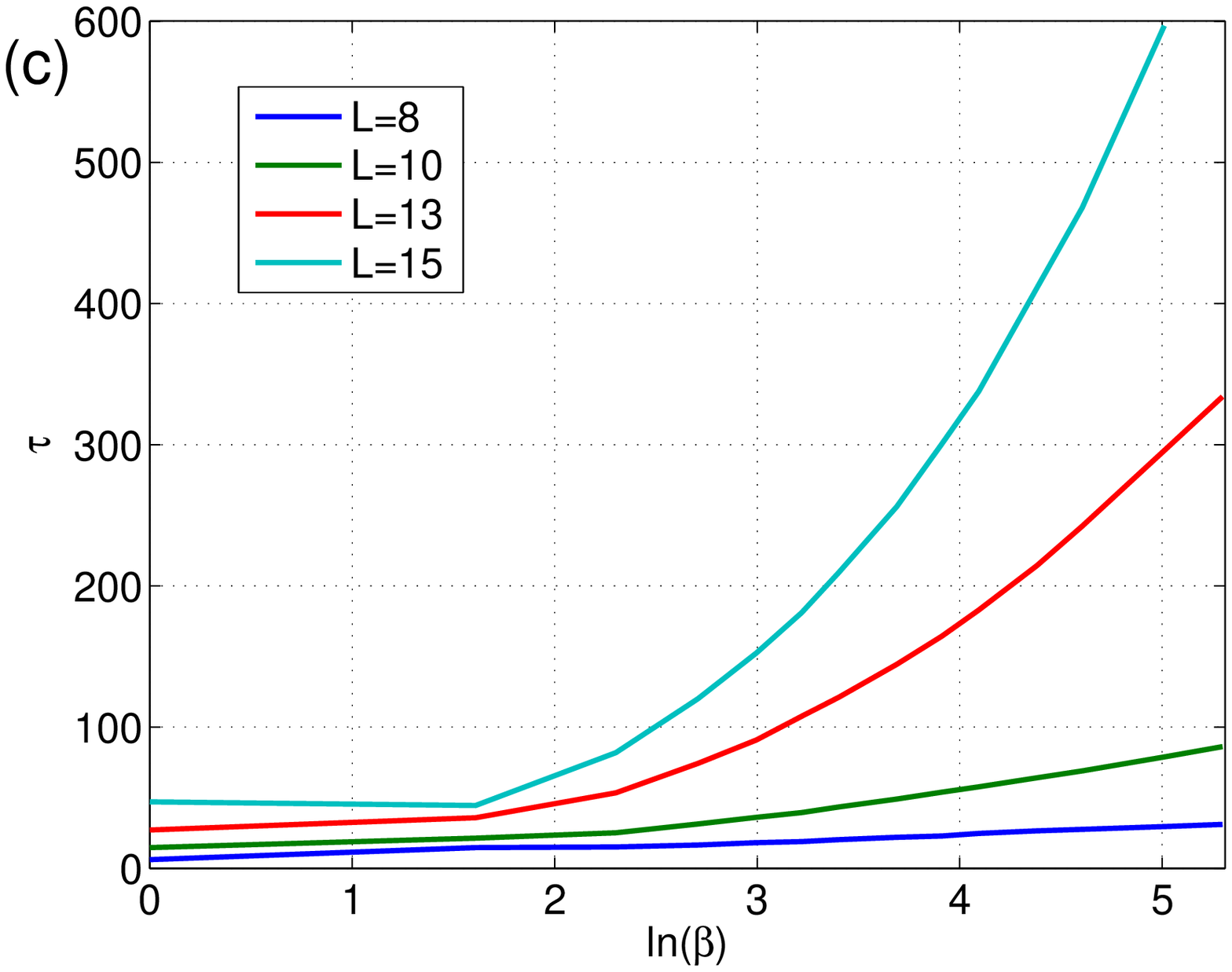}
\includegraphics[width=6cm,clip]{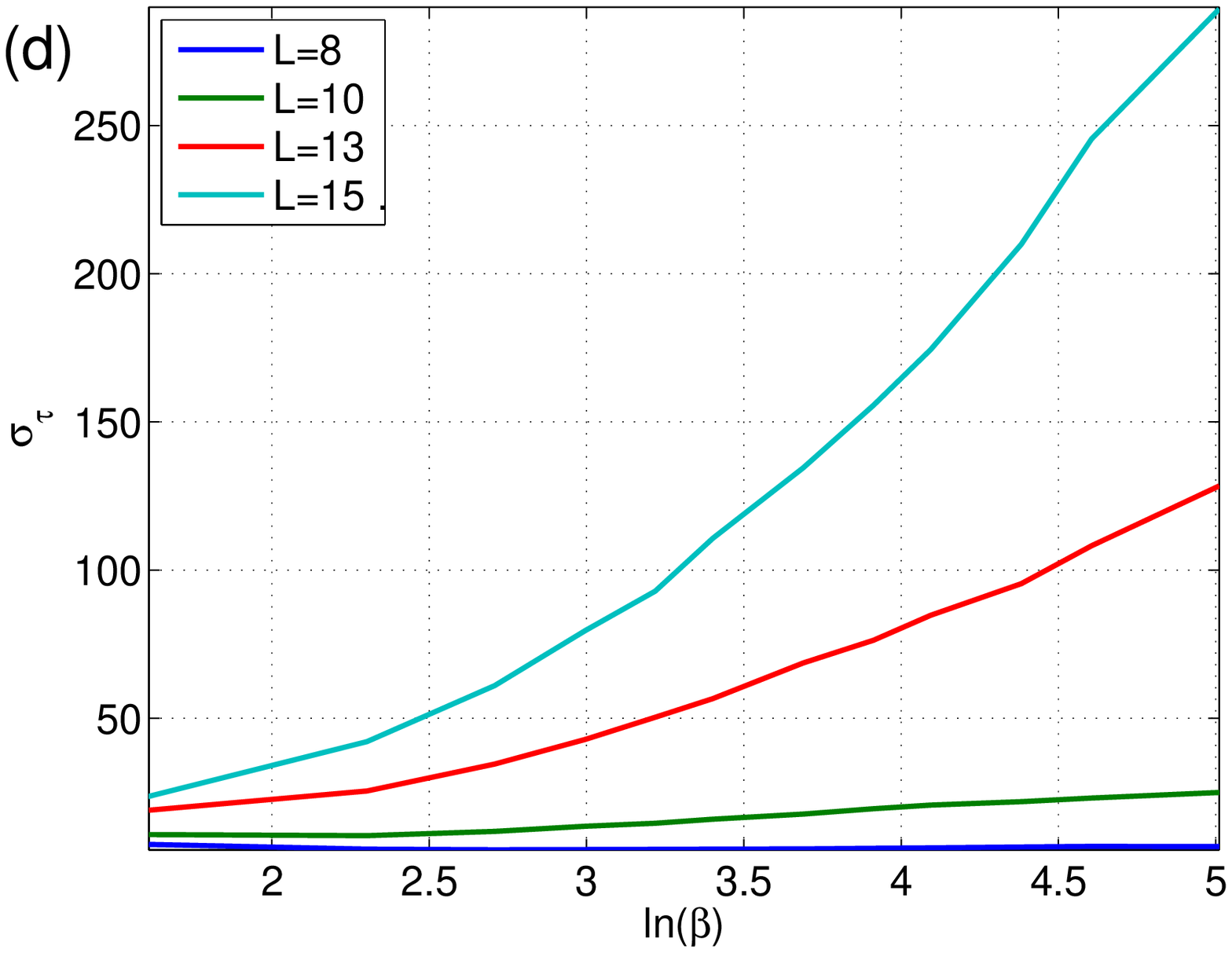}}
\caption{(a): Distributions of normalised durations of
reactive trajectories for $L=10$ in a range of $\beta$ compared to a Gumbel distribution.
(b): Average duration and variance of reactive trajectories as a function of $\ln(\beta)$ for $L=8$. (c): Average duration of reactive trajectories as a function of $\ln(\beta)$ for $8\le L\le 15$. (d) : Variance of duration of reactive trajectories as a function of $\ln(\beta)$ for $8\le L\le 15$.}
\label{tps}
\end{figure}

\begin{figure}
\centerline{\includegraphics[width=6cm]{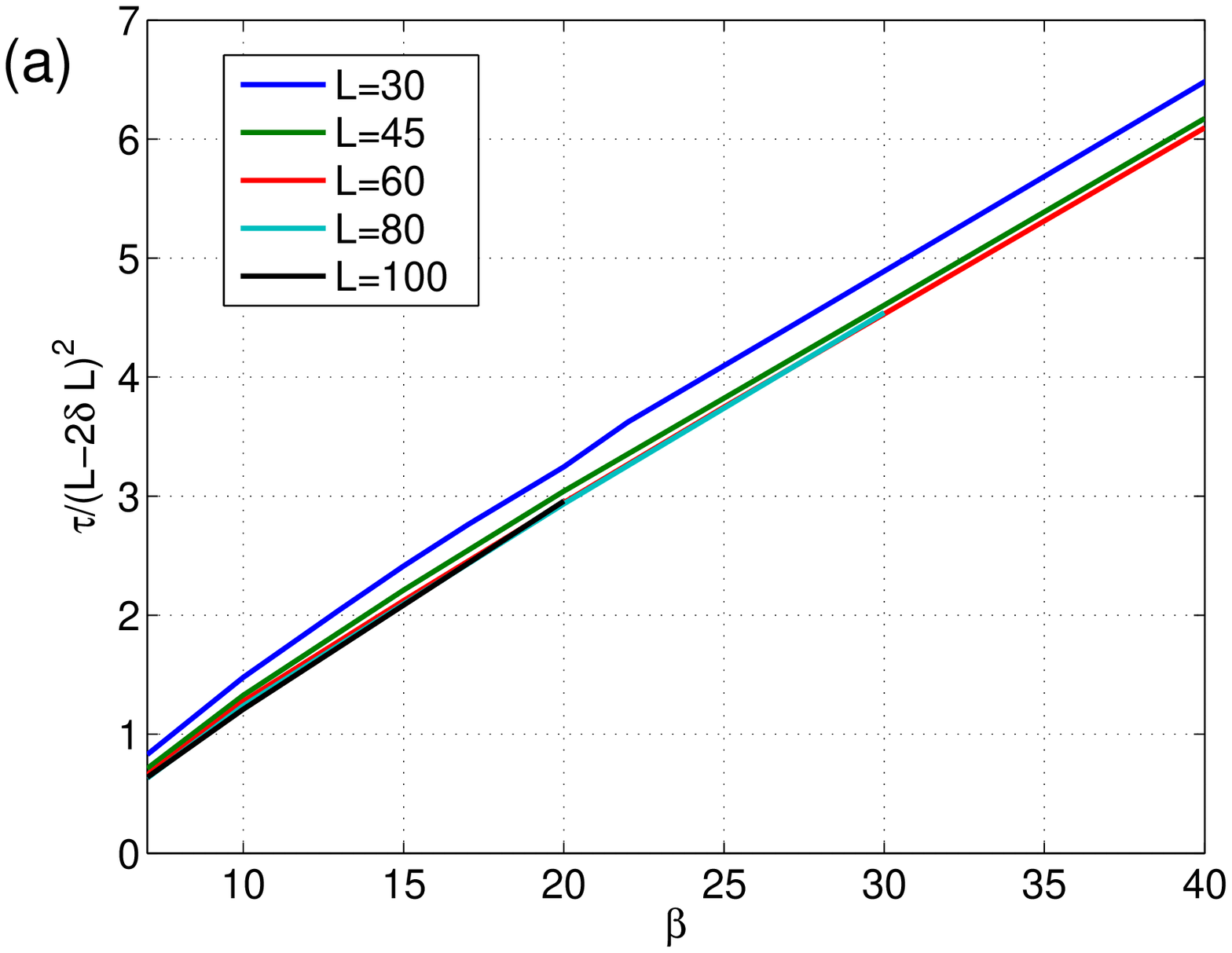}\includegraphics[width=6cm]{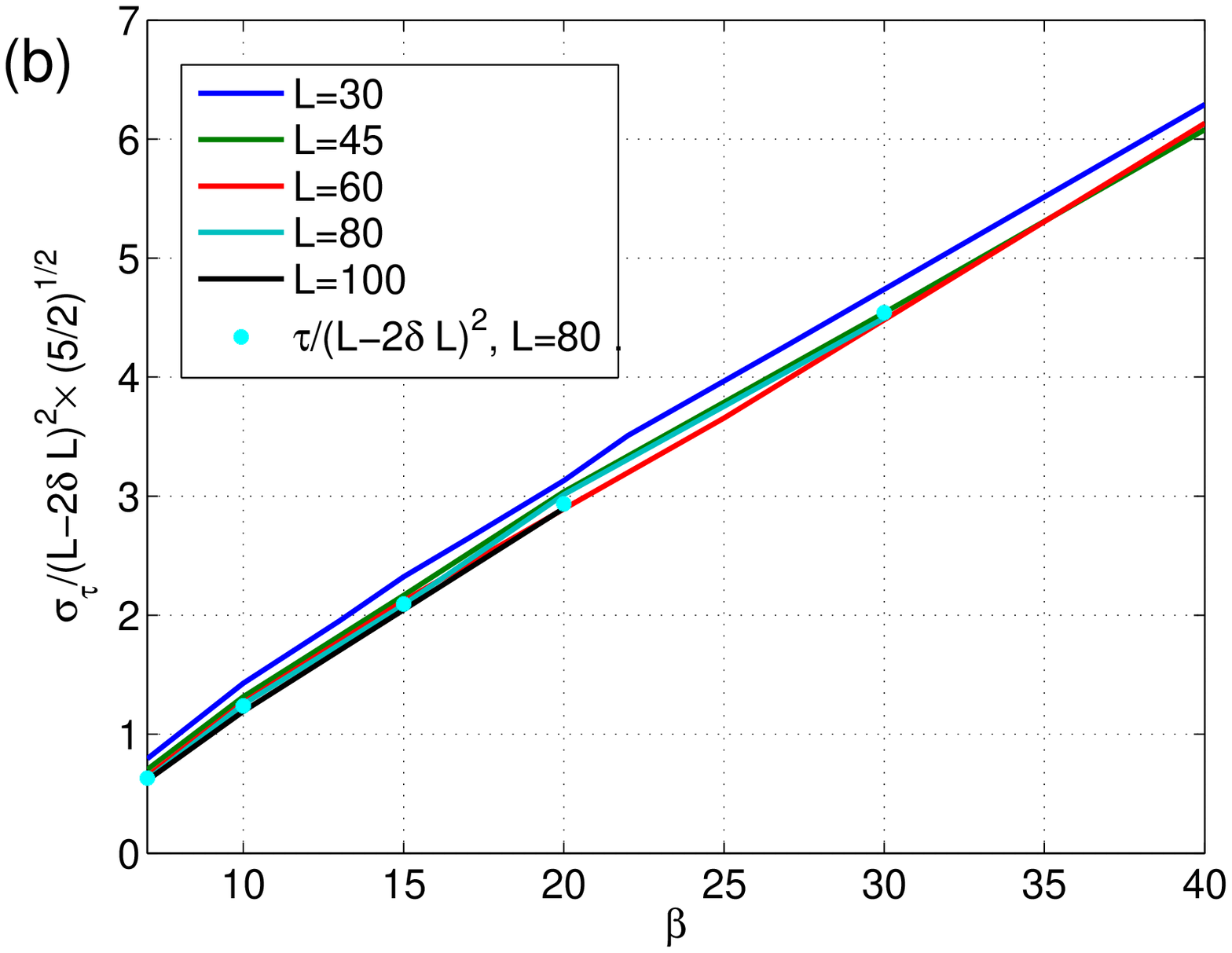}
}
\caption{(a): Average duration of the reactive trajectories rescaled by
$(L-2\delta L)^2$ as a function of $\beta$ for domain sizes larger than $30$. (b) Variance of the distribution of reactive trajectories durations rescaled by $\sqrt{5/2}$ and $(L-2\delta L)^2$ as a function of $\beta$ for domain sizes larger than $30$. The additional blue dots are an example of rescaled average duration.}
\label{tpsL2}
\end{figure}

We now investigate the statistical properties of the duration of reactive trajectories. As explained in section~\ref{sdur} the duration of a reactive trajectory is the time elapsed between the instant where the trajectory leaves $\mathcal{C}$ and the instant where it reaches $\partial \mathcal{B}$ (Fig.~\ref{sketch}). A realisation of the AMS with $N$ trajectories will thus give us $N$ durations. From theses $N$ durations, we can construct an empirical distribution of the trajectory durations (Fig.~\ref{tps}), estimate the mean duration $\tau$ (Fig.~\ref{tps} (b,c), Fig.~\ref{tpsL2} (a)) as well as the variance of the distribution (Fig.~\ref{tps} (b,d), Fig.~\ref{tpsL2} (b)). In practice, the quality of the sampling of the empirical PDF as well as the estimation of the mean and variance depends on $N$. One can verify numerically that we have central limit type interval of confidence on the cumulants \cite{jcp}. We chose $N=10000$ in the systematic study so that this interval of confidence is small enough.

We first focus on the non diffusive regime, at large $\beta$ and small $L$. We compare the numerical results to the analytical results for transition over a saddle in a one of freedom system discussed in section~\ref{sdur}. Figure~\ref{tps} (a) shows that the distribution of normalized durations is independent of $\beta$.
The comparison with a normalized Gumbel distribution (\ref{ngumb}) is relatively good good, in particular for $t\ge \tau$. We notice a slight systematic difference for $t\le \tau$. In order to go further with the comparison with the one degree of freedom case, we displays the average duration of trajectories and the variance of the distribution as a function of $\ln(\beta)$ in figure~\ref{tps} (b) for $L=8$. The data are consistent with $\tau$ being an affine function of $\ln(\beta)$ and  the variance of the distribution being nearly a constant, as in the one degree of freedom case. The estimate of the slope of $\tau(\beta)$ is $4.7$ while is variance is  $5.5\pm 0.2$. These values fall within $10$\% of the one degree of freedom prediction of a slope of $\tau(\beta)$ being $1/|\lambda_s|\simeq 5.3$ and a variance being $\pi/(|\lambda_s|\sqrt{6})\simeq 6.8$.

However, as the size $L$ is slightly increased, the average duration and the variance of the distribution departs much quicker from the asymptotic $\beta \rightarrow \infty$ regime than the mean first passage time. The average duration as a function of $\ln(\beta)$ is displayed in figure~\ref{tps} (c) for $8\le L\le 15$. One notice that for $L=15$, we are already leaving the regime of $\tau$ affine in $\ln(\beta)$. Moreover, the slope of $\tau$ Vs. $\ln(\beta)$ becomes clearly larger than $1/|\lambda_s|$ for $L=10$ and $L=13$. The departure is even more striking when we consider the variance of the distributions of durations as a function of $\ln(\beta)$ for $8\le L\le 15$ (Fig.~\ref{tps} (d)). While we do find the constant variance regime for $L=8$, we notice that for $L\ge 10$, we find a variance which is nearly affine in $\ln(\beta)$, so that ${\rm var}\propto \tau$. This situation is intermediate between the diffusive and the non diffusive regime (see \S~\ref{S3.3.3}). This illustrates the non-triviality of the question of the distribution of durations of reactive trajectories in stochastic partial differential equations.

We then move to the fully diffusive regime. It occurs for domain sizes larger than $L\ge30$ in the range of $\beta$ available to numerical simulations. Like in the non diffusive regime, we compute the average and variance of the duration of reactive trajectories (Fig.~\ref{tpsL2}). As explained in section~\ref{S3.3.3}, we expect trajectories to be random walks on the potential plateau and the trajectory durations to have the corresponding behaviour. Computing $\tau$ rescaled by $(L-2\delta L)^2$ as a function of $\beta$ for $30\le L\le 100$ strikingly confirm the result (Fig.~\ref{tpsL2} (a)): the values for $L\ge 45$ all fall on a master line. Note however that the slope is nearly $1/6$, as would be expected if the diffusion coefficient were $1/\beta$ and not $1/16$, as would be expected if the diffusion coefficient were $3/(8\beta)$ \cite{BBMP}. The discrepancy with the mathematical result remains to be explained. The variance of the trajectory durations is a growing function of $\beta$ which again corresponds very well to the random walk behaviour of the front (Fig.~\ref{tpsL2} (b)). Indeed, when rescaling it with $(L-2\delta L)^2$, it is linear in $\beta$. Moreover a second rescaling by $\sqrt{5/2}$ shows remarkable agreement with the average duration, indicating that ${\rm var}/\tau$ has the expected ratio.

\subsection{Escape probability}\label{S4.3}

We now consider estimates of the escape probability $\alpha$ (see also Appendix~\ref{Ad}). This is the probability to reach the set $\mathcal{B}$ before the set $\mathcal{A}$ starting from the iso surface $\mathcal{C}$ (see Fig.~\ref{sketch} (a)). It can formally be written with $P$ (see Eq.~\ref{Onsager-Machlup}, \S~\ref{S3.2}) by integrating over all the initial conditions $u_0$ on $\mathcal{C}$ and all the final conditions $u_1$ on $\partial \mathcal{B}$ \cite{ht}. Unlike the transition rate computed in the previous subsection, this quantity depends on the definition of the sets $\mathcal{A}$, $\mathcal{B}$ and $\mathcal{C}$. The escape probability has less physical importance than $1/T$. However, $\alpha$ corresponds to the committor (see \cite{tpt,ons}) whose asymptotic computation is difficult in many degrees of freedom systems. Since the committor is the solution of the same type of backward Fokker--Plank equation as the mean first passage time, large deviations results exist showing that it should decay exponentially like $\exp(-\beta \Delta V)$ \cite{ht,BEGK}, with $\Delta V$ the potential difference between the starting point on $\mathcal{C}$ and the lowest saddle. Moreover, this quantity is of fundamental numerical importance, as it is required to compute the transition rate. Besides, it is the only quantity on which mathematical results on the convergence of the algorithm exist (see \cite{cg07,PDM,STCO,poi,jcp,pdm2} and the appendix).

\begin{figure}
\centerline{
\includegraphics[width=12cm]{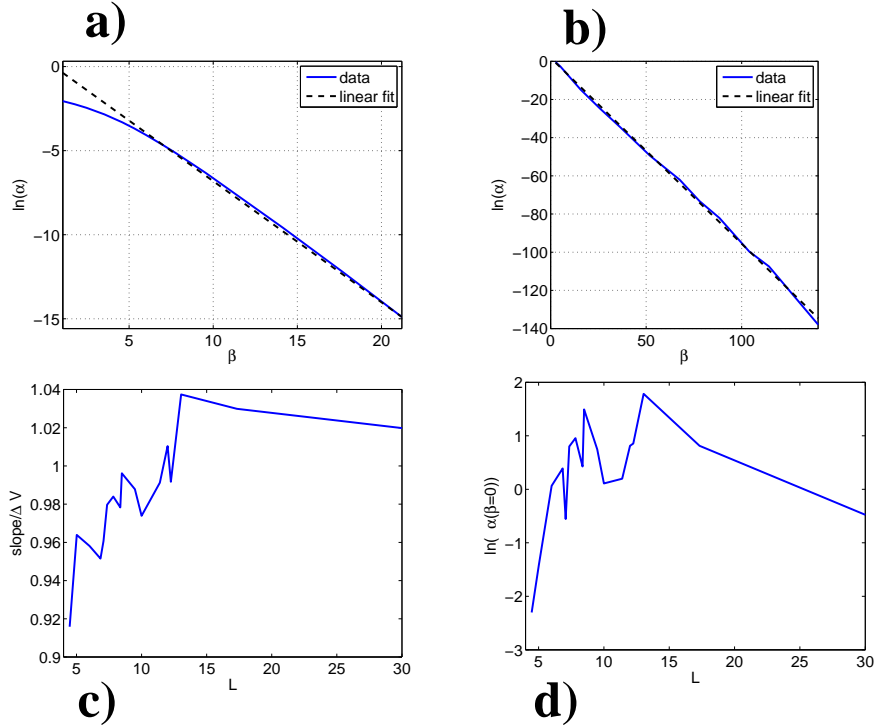}}
\caption{Examples of $\ln \alpha$ as a function of $\beta$ with the corresponding linear fit:
(a): $L=7.07$. (b): $L=17.3$. (c): $\frac{\ln \alpha}{\Delta V}$ as a function of $L$. (d):
Ordinate at the origin of $\ln(\alpha)$ at $\beta=0$ as a function of $L$.}
\label{res_AMS}
\end{figure}

We study how $\alpha$ depends on $\beta$
for sizes $L \in [4.5,30]$. The number of trajectories is $N=20~000$ and the number of
independent realisations of the algorithm is taken equal to 20.
Figure \ref{res_AMS} (a) clearly shows a change of regime with at $\beta \simeq 8$
. Above this value, we observe a large-deviation result where $\ln \alpha$
decreases linearly with $\beta$. We investigate more in details this behavior by dividing
the corresponding slope by $\Delta V =
V(A_{\rm saddle})-V(\mathcal{C})\simeq V(A_{\rm saddle})-V(A_0^\pm)$ using either $A_{\rm saddle} = A_0$ or $A_{\rm saddle} = A_1$
depending on the values of $L$.
This is shown in figure \ref{res_AMS} (c). The quantities are very close to the value +1
at $\pm 5\%$ precision which agrees very well with the large-deviation
predictions in these large $\beta$ regimes.
Note that these results are obtained for $\beta \gtrsim \beta^\star$ nearly independently of $L$.
A large deviation principle exists for the escape probability (Fig.~\ref{res_AMS} (a,b)), while
reactive trajectories are no more instantons (see Figs.\ref{possol}c and \ref{figinst}b).
Once again, it means that the escape probability is essentially related to the cost
for creating one front and it is a very robust mechanism \cite{VE}.

A significant advantage of AMS algorithm is that it gives precise information on Eyring--Kramers
prefactors. They can be obtained by computing the ordinate at the origin of $\ln \alpha$
as a function of the size $L$ for instance. This is shown in Fig. \ref{res_AMS} (d) with
prefactors belonging to $~[0.1,10]$. The variation with size of the ordinate at origin is likely related to
the numerical precision on the slope: an overestimate of the slope in absolute value leads to
an overestimate of the ordinate and vice versa. However at fixed $\beta$, one generally notes a decrease of $\alpha$ with $L$.

\subsection{Phase diagram}
\begin{figure}
\centerline{
\includegraphics[width=10cm]{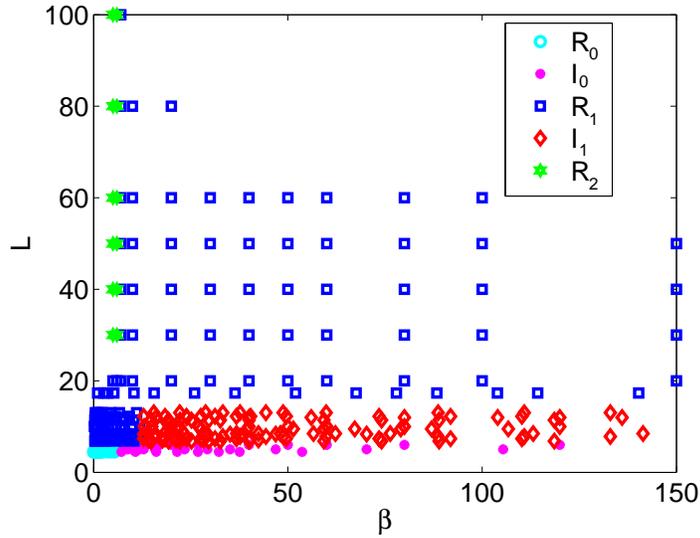}
}
\caption{Phase diagram of possible reactive trajectories in the $(\beta,L)$ plane using AMS
algorithm to be compared with figure~\ref{diag1}. In particular, we use the same notations: $I_0$: instanton flip, $I_1$: instanton with one front, $R_0$: random walk flip, $R_1$: random walk with one front, $R_2$: random walk with two fronts.}
\label{diag2}
\end{figure}

We give a synthesis of these results by constructing a numerical version of the diagram
shown in Fig. \ref{diag1}. One can determine the number of fronts through visualisation of
the space-time representations. The discrimination between a flip and a one-front trajectory
is more easily done by looking at the average duration (see Fig. \ref{tps}c).
However, visualisation alone is not enough to reasonably determine whether a trajectory is an instanton or a
random-walk trajectory. For a given point $(\beta,L)$ in parameter space, we consider a trajectory
to be an instanton if the average duration and the mean first passage time falls into the non-diffusive regime (see Fig. \ref{tps}c Vs. Fig.~\ref{tpsL2} and Fig.~\ref{mfpt_fig}). For most sizes, the transitions belong clearly to one regime or another. We assumed that the transition was out of the low noise regime if the relative difference between the mean first passage time calculated using AMS and using the Eyring--Kramers formula was more that $15$\%. The precise position of the boundary between regime will depend on the value of this tolerance.
This criterion gives a lower bound $\beta \gtrsim 12$ and an upper bound $L\lesssim 13$ for the instanton regime.
All these informations are encoded in the phase diagram of figure~\ref{diag2}. There is a very good agreement with
the theoretical predictions both on the different regions of parameter space and the type
of reactive trajectories (Fig.~\ref{diag1}). In our simulations, we were not able to obtain reactive
trajectories with more than two fronts. The exponential curves at the boundary of
some of the regions are not resolved here and appears to behave as straight lines, as a
consequence of the rather poor grid resolution of the $(\beta,L)$ plane.

\section{Conclusion}
\label{S5}
This article intends to present a complete theoretical and numerical analysis
of transition rates and reactive trajectories between two metastable states of the 1-D Allen--Cahn equation.
In this work, we have used a recent algorithm called adaptive multilevel splitting
dedicated to the computation of very-low probability events. We are able
to numerically investigate metastability of the Allen--Cahn equation in  a fast and efficient way
by systematically computing reactive trajectories using this algorithm.
This is one of the main aspect of this work.
It allows us to synthesise existing asymptotic theories like
Freidlin--Wentzell large deviations \cite{ht}, Eyring--Kramers law of mean first passage time
\cite{ha,Kramers,Eyring,JSL}, and theory of front motion \cite{wall_pre,BBMP}.
The numerical results validate these theories but also bridge the gap between them and go further.

The combined use of numerics and theory shows that in the zero noise limit, reactive
trajectories are either global reversal of the flow when the domain size $L$ is small,
or a front nucleation at one end of the domain and its propagation toward the other end
provided $L$ is sufficiently large. Solutions with several fronts
are not global minima of the action \cite{VE,wall_pre}, and are never observed numerically for large enough $\beta$.
We are also able to compute with good accuracy both the escape probability and the
transition rate in many different regimes.
We show that the escape probability has a very robust large-deviation behavior in
the limit of zero noise, as expected
from Freidlin--Wentzell theory. Moreover, in regimes where Eyring--Kramers law
does not apply anymore, we always observe a large-deviation behavior of the transition rate whenever the escape probability does. It calls for possible further investigations
on the prefactor alone in these situations.
Another remarkable result is the robustness of the large-deviation behavior
of the escape probability at finite-amplitude noise
and when the most probable trajectories are no more instantons. Theoretical predictions, especially those on size dependence, are in fact valid in a much larger range of parameters than the low noise limit $\beta \rightarrow \infty$.
The validity of the prediction merely requires the creation of a single front and a long
enough reactive trajectory \cite{VE}.

The AMS algorithm is also able to provide accurate statistics on the average duration $\tau$
of reactive trajectories for which no theory exists yet. Remarkably, we
find similar scaling laws in Allen--Cahn equation
than for systems with only one degree of freedom, for which theoretical analysis exist \cite{asy}.
We obtain that $\tau \propto \ln \beta \times {\rm exp}(L/\sqrt{2})$ when the curvature at the saddle
is large, and $\tau \propto \beta L^2$ when it is small.

This work supports the idea that AMS is a very efficient tool for analysing extremely rare transitions
in systems with a large number of degree of freedom and SPDEs as well. It is a promising approach
for studying more realistic systems arising from practical problems \cite{Ey,kuro}.
Although Allen--Cahn equation has a rather
simple structure as being a gradient system, AMS is likely to give useful insights for
non-gradient systems as well, as there are no principle limitations in its use.
In particular, an interesting step is to consider 2-D non-gradient SPDEs such as the stochastic
2-D Navier--Stokes equations \cite{sns} whose theoretical description is much more subtle
than for gradient systems \cite{bt_niso}.
We are confident that AMS approach will becomes more popular in the future due to its
simplicity, robustness and efficiency.

\appendix

\section{Appendices}
We first develop the calculation of the approximation of the negative eigenvalue of the Hessian of the potential $V$
We then give here some details on the AMS algorithm, first we provide a detailed description
of the algorithm itself, then some basic applications like the computation of mean first
passage time. We end with a practical discussion on some numerical convergence issues.

\subsection{Approximating the Eigenvalue $\lambda_s$}
\label{Ap_exponential_interactions}
In this first appendix, we detail the calculation of an approximation of the first eigenvalue $\lambda_s$ of the Hessian of $V$, presented in section~\ref{exponential_interactions}. This eigenvalue controls the prefactor of the mean first passage time (Eq.~\ref{eqpref}).

The eigenvalues $\lambda_i$ and eigenmodes $\Phi_i$ are the solution of the problem
\begin{equation}
\underbrace{ -\partial_x^2 \Phi_i + (3A_{s,0}^2-1)}_{H_{s,0}}\Phi_i=\lambda_{i,A_{s,0}} \Phi_{i,A_{s,0}}\,,
\label{eigvalprob}
\end{equation}
 with boundary conditions $\Phi_i(0)=\Phi_i(L)=0$. The eigenvalue problem can be posed either at the saddle, for $A_s$, or in the minimum $A_{0}$. The operator $H_{s,0}=\nabla^2 V|_{A_{s,0}}$ is the hessian of the potential $V$. It is self adjoint on the set of function (regular enough) that cancel out on both ends of the domain $x=0,L$. It is well known that in the limit of infinite size $L\rightarrow \infty$ (see \cite{wall_pre} and references within): we have
\begin{equation}
A_s=A_{1,\infty}=\tanh\left( \frac{x-\frac{L}{2}}{\sqrt{2}}\right)\,,\,
\Phi_\infty=1-\tanh^2\left( \frac{x-\frac{L}{2}}{\sqrt{2}}\right)\,,\,
\lambda_\infty=0\,.
\end{equation}
This eigenmode is a goldstone mode, related to the translational invariance of the front in an infinite domain \cite{jzj}. We denote hyperbolic tangent $\mathcal{H}_{u}(x)\equiv\tanh((x-u)/\sqrt{2})$, which will simplify our notations.

How $\lambda_s$ converges toward $0$ in a finite domain of size $L$ ? In order to address that question, we will provide an upper bound for $\lambda_s$. For that matter, we use the fact that $H_s$ is self-adjoint. Its Rayleigh quotient
\begin{equation}
R(H_s,f)=\frac{\int_0^L fH_s(f)\, {\rm d}x}{\int_0^L f^2 \, {\rm d}x}\,,\, f(0)=f(L)=0\,,
\end{equation}
 then verifies the min-max theorem, which states that $R(H_s,f)$ is bounded by the lowest ($\lambda_s$) and highest eigenvalue of $H_s$. Moreover, we have that $R(H_s,\Phi)=\lambda_s$ and that given $\Psi$ an approximation of $\Phi$, $R(H_s,\Psi)$ will converge (from above) toward $\lambda_s$ as $\Psi$ converges toward $\Phi$.

\begin{figure}
\centerline{\includegraphics[width=5.5cm]{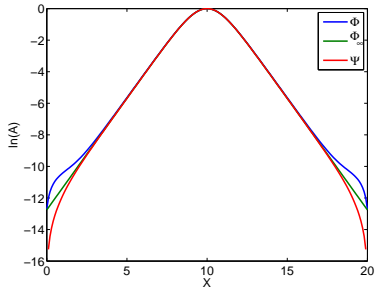}}
\caption{Logarithm of the eigenmode and its approximations in a domain of size $L=20$.}\label{figapprox2}
\end{figure}

We will now tailor an analytical approximation of $\Phi$ and $A_1$ so as to give an upper bound on $\lambda_s$. This upper bound will serve as an approximation.  The lowest eigenvalue $\lambda_s$ and the eigenmode $\Phi$ correspond to the unstable direction of $V$ at the saddle: small motion of the position of the front near $L/2$. If the field $A$ is slightly displaced away from the saddle $A_1$ along the unstable direction, the front (located at position $y$) is displaced by a small $\delta x=y-L/2$. We then have $A\approx A_1+\delta x\Phi$. In order to obtain an approximation of $\Phi$, we start from an analytical approximation $K(x,y)$ of the field $A(x)$ when a front is located at position $y$
\begin{equation}
K(x,y)\equiv \mathcal{H}_0(x) \mathcal{H}_L(x) \mathcal{H}_y(x)\,.\label{ansatzfront}
\end{equation}
Performing the expansion gives us
\begin{equation}
\approx K\left(x, \frac{L}{2} +\delta x\right)=K\left(x,\frac{L}{2} \right)+\delta x \frac{\partial K}{\partial y}\left(x,\frac{L}{2} \right)+O(\delta x^2)\,.
\end{equation}
This leads us to choose
\begin{equation}
\Psi\equiv-\sqrt{2}\frac{\partial K}{\partial y}\left(x,\frac{L}{2} \right)=-\mathcal{H}_0(x) \mathcal{H}_{L}(x)\left(1- \mathcal{H}_{\frac{L}{2}}(x)^2\right)\,.
\end{equation}
The minus sign and the square root of $2$ simply ensure that $\Psi$ is a positive function whose maximum is $1$ at $x=L/2$. In order to give a good analytical approximation of the ratio of integrals $I=R(H_s,\Psi)$, we set
\begin{equation}
\tilde{A}_1\equiv\mathcal{H}_0(x)+\mathcal{H}_{L}(x)-\mathcal{H}_{\frac{L}{2}}(x)
\,,\, \tilde{I}\equiv\frac{\int_0^L \Phi_\infty \left(-\partial_x^2  + (3\tilde{A}_{1}^2-1) \right)\Phi_\infty\, {\rm d}x}{\int_0^L \Phi_\infty^2 \, {\rm d}x}\,.
\end{equation}
Note that $\Phi_\infty$ does not cancel out at $x=0,L$, nor is $\tilde{A}_1$ strictly the saddle point of $V$. As a consequence $\tilde{I}$ is not strictly speaking a Rayleigh quotient, nor should we \emph{a priori} expect to verify exactly its properties. The ratio $\tilde{I}$ is a good approximation of $I$ because $\Phi_\infty^2$ like $\Psi^2$ decrease like $1/\cosh^4$ away from $L/2$, up to corrections on the boundaries (Fig.~\ref{figapprox2}). Any relative contribution to the integrals near $x=0$ and $x=L$, where the errors are made, are exponentially small by a factor $\exp^4(-L/(2\sqrt{2}))=\exp(-L\sqrt{2})$ relatively to the contributions from the neighbourhood of $x=L/2$. Meanwhile, the relative difference between $\Phi_\infty$ and $\Psi$, or between $\tilde{A}_1$ and $A_1$ are also exponentially small near $L/2$. Note however, that it is fundamental that $A_1$ has the proper boundary conditions at $x=0$ and $x=L$, as it is they that will contribute to the exponential decay of the eigenvalue. The quality of this approximation depends on how large $L$ is. For $L=10$, $\exp(-L\sqrt{2})\simeq 7\cdot 10^{-7}$: the system is quickly in the range of validity.

The denominator is calculated in the same way as $\Delta V_1$, it is equal to $(4\sqrt{2})/3$. In order to calculate the numerator, denoted $J$, we first note that
\begin{equation}
\tilde{H}\Phi_\infty=-3\left(1-\mathcal{H}_{\frac{L}{2}}(x)^2\right)\left(\underbrace{\mathcal{H}_0(x)+\mathcal{H}_L(x) }_{\equiv S}\right) \left(-\underbrace{\left(\mathcal{H}_0(x)+\mathcal{H}_L(x)\right)}_{=S} +2\mathcal{H}_{\frac{L}{2}}(x) \right)\,.
\end{equation}
We process the sum
\begin{equation}S=2e^{-\frac{L}{\sqrt{2}}}\frac{\exp\left(\frac{2x-L}{\sqrt{2}}\right)-\exp\left(-\frac{2x-L}{\sqrt{2}}\right)}{1+\exp\left(\sqrt{2}(x-L)\right)+\exp\left(-\sqrt{2}x\right)+\exp\left(-\sqrt{2}L\right)}\,.\end{equation}
 In the neighbourhood of $L/2$, only $1$ has a significant contribution in the denominator, so that this sum can be approximated by the numerator in $\tilde{H}\Phi_\infty$. Note that it is negligible compared to $\tanh((x-L/2)/\sqrt{2})$: Both cancel out in $L/2$, while only the hyperbolic tangent is of order $1$ in the direct neighbourhood. Now $J$ reads
\begin{equation}J=-12e^{-\frac{L}{\sqrt{2}}}\int_{0}^L \left(1-\mathcal{H}_{\frac{L}{2}}(x)^2\right)^2\mathcal{H}_{\frac{L}{2}}(x)\left(\exp\left(\frac{2x-L}{\sqrt{2}}\right)-\exp\left(-\frac{2x-L}{\sqrt{2}}\right) \right) \, {\rm d}x\end{equation}
We eventually note that the only contributions to the integral are around $L/2$, so that we can let the bounds go to infinity. This gives us $J=-32\sqrt{2}e^{-L/\sqrt{2}}$, then
\begin{equation}
\tilde{I}=-24e^{-\frac{L}{\sqrt{2}}}\,. \label{l1_bis}
\end{equation}
This is an approximation at first order in $\exp(-L/\sqrt{2})$: the corrections are exponentially small as $L$ takes large values that can however be considered in AMS simulations. A numerical calculation of the integral in $I$ confirms this assertion.

\subsection{Algorithm description \label{Ad}}
The general setting is to have a Markov process describing our system, i.e.
$(X_t^{x_0})_{t \geq 0}$ with $X_0^{x_0} = x_0 \in {\cal E}$.
The phase space ${\cal E}$ of the system can be either finite-dimensional of infinite-dimensional.
We assume that there are two sets ${\cal A}$ and ${\cal B}$ with ${\cal A} \cap {\cal B}
= \emptyset$. The goal is
to estimate the probability $\alpha$
to reach ${\cal B}$ starting from the initial condition $x_0$ before
going to ${\cal A}$. If $\tau_{\cal C}$ denote the stopping time defined as
$\tau_{\cal C}(x) = \inf \{ t \geq 0, X_t^x \in {\cal C}, \}$ for
a given Borel set ${\cal C}$, the problem translates into computing
\begin{equation}\label{pb}
\alpha \equiv \alpha(x_0) = \Pr(\tau_{\cal B}(x_0) < \tau_{\cal A}(x_0)).
\end{equation}
A reactive trajectory is a particular realisation of (\ref{pb}). Note that this probability,
when seen as a function of $x_0$, takes values zero if
$x_0 \in {\cal A}$ and one if $x_0 \in {\cal B}$. It is called
{\it committor} or {\it importance function} or {\it equilibrium potential} in mathematics.
In particular, for diffusive processes, it solves the backward Fokker--Planck equation
with boundary conditions $\alpha(x)=0, x\in {\cal A}$ and $\alpha(x)=1,x\in {\cal B}$. The algorithm not only
gives a pointwise estimate of
the backward Fokker--Planck but also provides the ensemble associated with the
probability. It is clear that for high-dimensional systems, access to the Fokker--Planck equation
or doing direct Monte-Carlo simulations when $\alpha$
is too small is, most of the time, out of question.

The main idea is to introduce a scalar quantity  which measures how far a trajectory is escaping
from the set ${\cal A}$ before returning to ${\cal A}$. There is no unique way to measure
how far a trajectory is escaping from the return set ${\cal A}$, or put in another way there is
no unique order relation in the space of trajectories.
However, a natural choice is to consider the following quantity
\begin{equation}\label{Qx}
Q(x) = \sup_{t \in [0,\tau_{\cal A}]} \Phi(X_t^x),~ \Phi: \field{E} \to \field{R}.
\end{equation}
The function $\Phi$ which is often renormalized to be in $[0,1]$ is called {\it reaction
coordinate} or {\it observable} and is such that $\Phi({\cal A}) = 0$ and $\Phi({\cal B}) = 1$.
Note that the choice of $\Phi$ might become crucial and we will comment it hereafter.
The algorithm is doing the following (see also Fig. \ref{sketch}):
\\\\
{\bf Pseudo-code}. Let $N$ be a fixed number, and $\Phi$ given.
\begin{itemize}
\item[]{\it \underline{Initialization}}.\\
Set the counter ${\cal K} = 0$.
Draw $N$ i.i.d. trajectories indexed by $1 \leq i \leq N$, all starting at $x_0$ until
they either reach ${\cal A}$ or ${\cal B}$.
Compute $Q_i(x_0)$ for $1 \leq i \leq N$. In the following we will write
$Q_i$ these quantities since their initial restarting conditions might change.
\\\\
DO WHILE $\left({\displaystyle \min_i  Q_i  \leq b}\right)$

\begin{itemize}
\item[] {\it \underline{Selection}}: find
$$
i^\star = argmin_i ~Q_i.
$$
Let $\widehat Q \equiv Q_{i^\star}$.\\
\item[] {\it \underline{Branching}}:
select an index $I$ uniformly in \\$\{1,\cdots N\} \setminus \{i^\star\}$.
Find $t^\star$ such that
$$
t^\star = \inf_{t} \{ \Phi\left(\{ X_t \}_{I}\right) \geq \widehat Q \}
$$
Compute the new trajectory $i^\star$ with initial condition $x^\star \equiv \{X_{t^\star} \}_I$
until it reaches either ${\cal A}$ or ${\cal B}$.\\
Compute the new value $Q_{i^\star}(x^\star)$.
$$
{\cal K} \leftarrow {\cal K} + 1
$$
\end{itemize}
END WHILE
\end{itemize}
The performance of the algorithm depends rather crucially on the choice of $\Phi$.
In fact, it is possible to show that when $\Phi(x) = \alpha(x)$ (see eq.(\ref{pb})) or
when the system has only one degree of freedom (${\cal E} = \field{R}$),
the number of iterations ${\cal K}$ when the algorithm has stopped
has a Poissonian distribution \cite{poi,STCO}
\begin{equation}
{\cal K} \sim {\rm Poisson}(-N \ln \alpha).
\end{equation}
In other words, the quantity $\frac{K}{N}$ yields an estimate of $-\ln \alpha$
(see \cite{jcp} for particular SDEs).
Moreover,
\begin{equation}
\widehat \alpha = \left(1 - \frac{1}{N}\right)^{\cal K}
\end{equation}
gives an estimate of $\alpha$ itself. In some cases, one can demonstrate a central limit theorem for $\widehat \alpha$, with a variance scaling like $1/\sqrt{N}$
\begin{equation}
\sqrt{N}\left(\frac{\alpha-\widehat{\alpha}}{\alpha}\right)\xrightarrow[N\rightarrow \infty]{\mathcal{D}}\mathcal{N}(0,|ln(\alpha)|)\,.
\end{equation}
Some sets of hypotheses lead to a bias behaving as $\frac{1}{N}$ \cite{cg07,pdm2}. With more loose hypotheses or in more complex cases, the variance still scales like $1/\sqrt{N}$, but can be larger than what is predicted \cite{jcp}.

Note that for arbitrary $\Phi$
and in dimension larger than 1, the statistical behavior of the algorithm is
a difficult mathematical question and it is still open \cite{STCO}. In particular,
it is unclear whether these estimates are biased or not, even in the limit of $N \to \infty$.
In practice, a ``bad" reaction coordinate which differs significantly from the optimal committor,
has a fat-tailed distribution although its variance still behaves like
$\frac{1}{\sqrt{N}}$. Therefore, the larger $N$, the better precision one obtains.
Overall, the algorithm numerically exhibits very robust results.
There are other formulation of the algorithm, in particular, it is possible to kill
a given proportion of the $N$ trajectories at each algorithmic step
(typically, if one kills $p$ trajectories at each step followed by replacement,
an estimate of the probability becomes $(1-\frac{p}{N})^{\cal K}$). This version has the advantage
to be computationally more efficient when parallelized (\cite{jcp}).

\subsection{Practical applications for Allen--Cahn equation}\label{Ac}

\subsubsection{Reaction coordinates}

In our case the set ${\cal A}$ is a neighborhood of the stationary solution  $A_0^+$
and the set ${\cal B}$ a neighborhood of $A_0^-$.
The choice of the reaction coordinate is often dictated by the problem itself, a natural
one is to choose a quantity which tells how far one is from $A_0^+$. We use the following
reaction coordinates
\begin{equation}\label{GLobs}
\Phi_l(A) = \frac12 \left( 1-\frac{\overline{A}}{\overline{A_0^+}} \right),~
\Phi_n(A) = \Phi_l(A) + \left|\frac{1}{\sqrt{L}} \frac{<A,A_1>}{||A_1||} \right|,
\end{equation}
where $\overline{A} = \int_0^L A(x)~dx$ and $<A,B> = \int_0^L A(x)B(x)dx, ||A||^2=\int_0^L A^2(x)dx$.
These reaction coordinates have the property that they vanish at $A_0^+$ and are equal to 1
at $A_0^-$.
The second reaction coordinate $\Phi_n$ gives some additional weights to the fronts present in $A$.
 In practice, we chose
\begin{equation}\partial \mathcal{A}=\{A \backslash \Phi_n(A)=0 \}\,,\, \partial \mathcal{B}=\{A \backslash \Phi_n(A)=1 \}\,,\,\mathcal{C}=\{A \backslash \Phi_n(A)=0.05 \}\,.\label{defabc}
\end{equation}
For domain of large size ($L=60$), both values $\Phi_n=0.05$ and $\Phi_n=0.025$ were tested. The two definitions led to differences in computed quantities ($\widehat{\alpha}$, $\tau$) clearly smaller than their variance. For this reason, we consider than this change of definition of $\mathcal{C}$ does not play any role on the final result. However, this change plays a clear role in the computation time of the generation of initial conditions distributed on $\mathcal{C}$, especially when $\beta$ or $L$ are large. For this reason, $\Phi_n=0.025$ was systematically used for computations at $L=80$ and $L=100$.

\subsubsection{Statistics from the output \label{sout}}

There are two important quantities which can be obtained from AMS output:
\begin{itemize}
\item[$\bullet$] {\it distribution of duration of reactive trajectories}. In \cite{jcp},
a numerical analysis shows the convergence of this distribution. In the case of SPDE and Allen--Cahn
for example, there is no theoretical results on this distribution. In our case, one
can check for instance the rate of growth of these lengths with $\beta$. More generally,
since AMS is providing an ensemble of reactive trajectories, it is possible to
perform many a-posteriori statistics which become accurate as $N$ is larger.
\item[$\bullet$] {\it mean first passage time} In order to compute this quantity, one defines
a isosurface ${\cal C}$ surrounding the set ${\cal A}$. One generates an ensemble of $N$
trajectories which all start at $A_0^+$. One then estimates the mean time $\tau_1$ it takes to
cross the surface ${\cal C}$ and $\tau_2$ the mean duration of nonreactive trajectories
(that is, the length of trajectories between the hitting time of ${\cal C}$ and the
return time to ${\cal A}$) \cite{cglp}. Assuming that there are a mean number $<n>$
of nonreactive trajectories, the mean first passage time $T$ is
\begin{equation}\label{mfpt_ams}
T = <n>(\tau_1+\tau_2) + (\tau_1+\tau_r),
\end{equation}
where $\tau_r$ the mean duration of reactive trajectories
found by the algorithm. The number of failed attempts $<n>$ is in fact related to $\alpha$ by
$\alpha = \frac{1}{<n> + 1}$.
\end{itemize}

\subsubsection{Convergence issues}\label{Aconv}
There are two sets of parameters to take into account: first, the SPDE discretisation with
timestep $dt$ and spatial resolution $dx$, second, the number $N$ of the trajectories in the
algorithm and the choice of the reaction coordinate $\Phi$.  We first illustrate the convergence
properties of the SPDE itself. We use a test case of $L=7.07$ using $N=20~000$ trajectories and
the norm reaction coordinate $\Phi_n$ (which yields better results than $\Phi_n$ \cite{jcp}).
We increase the spatial resolution from $dx=\frac16$ to $dx=\frac{1}{18}$ and decrease the
timestep $dt$ accordingly. Note that, in order to ensure the stability of the numerical
integration of the diffusive part, the constraint is to have $D > \frac{dt}{dx^2}$.
The choice of $dt$ in addition to this constraint crucially depends on the accuracy of
hitting times (of ${\cal A}$ and isosurfaces of the reaction coordinate $\Phi$)
needed in the algorithm. It is
well-known that these hitting times converge like
$\sqrt{dt}$ \cite{ght} bringing a limit to the accuracy of the solution and the probability
$\alpha$ found by the algorithm as well (see \cite{jcp}).

\begin{figure}
\centerline{
\includegraphics[width=6cm]{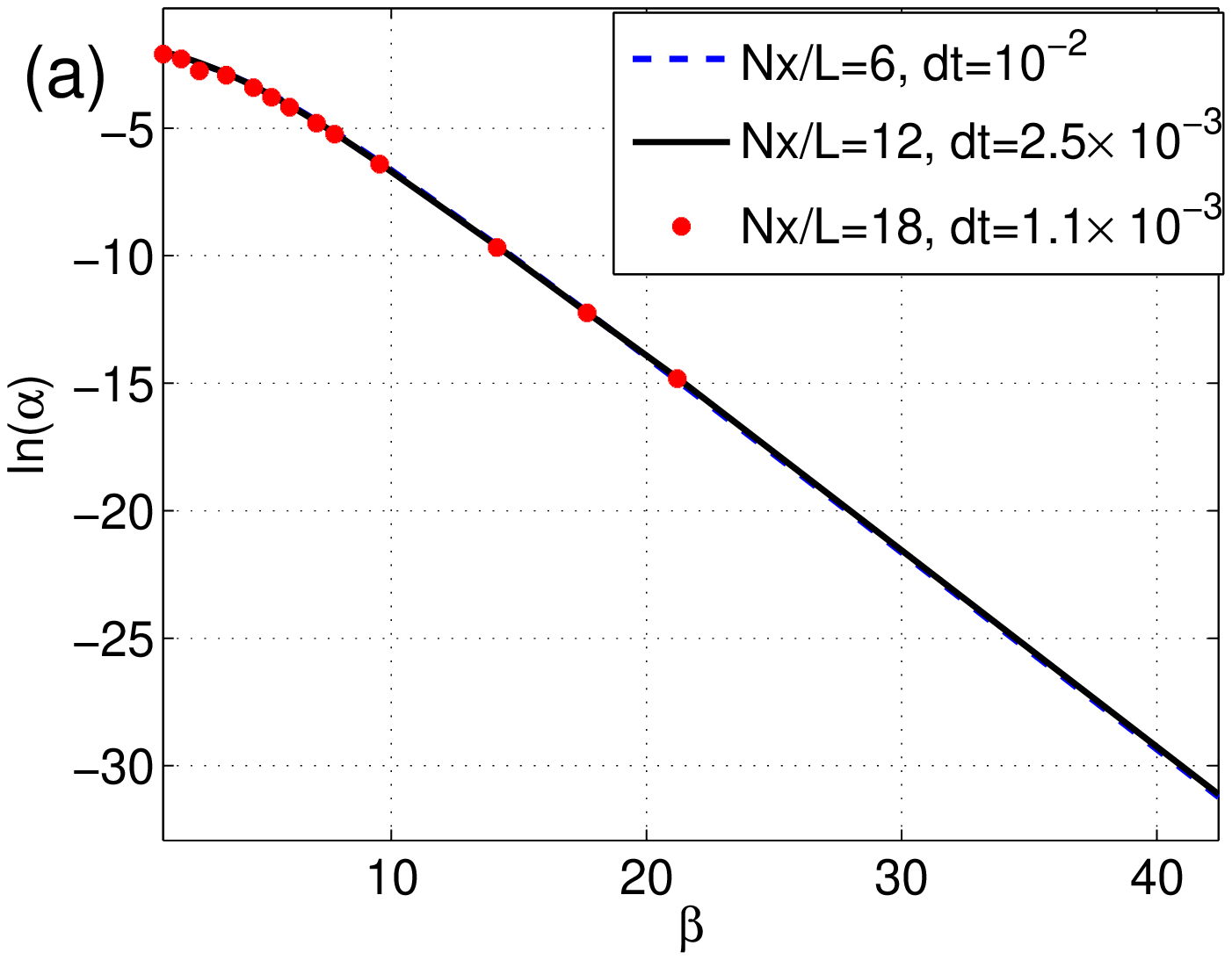}
\includegraphics[width=6cm]{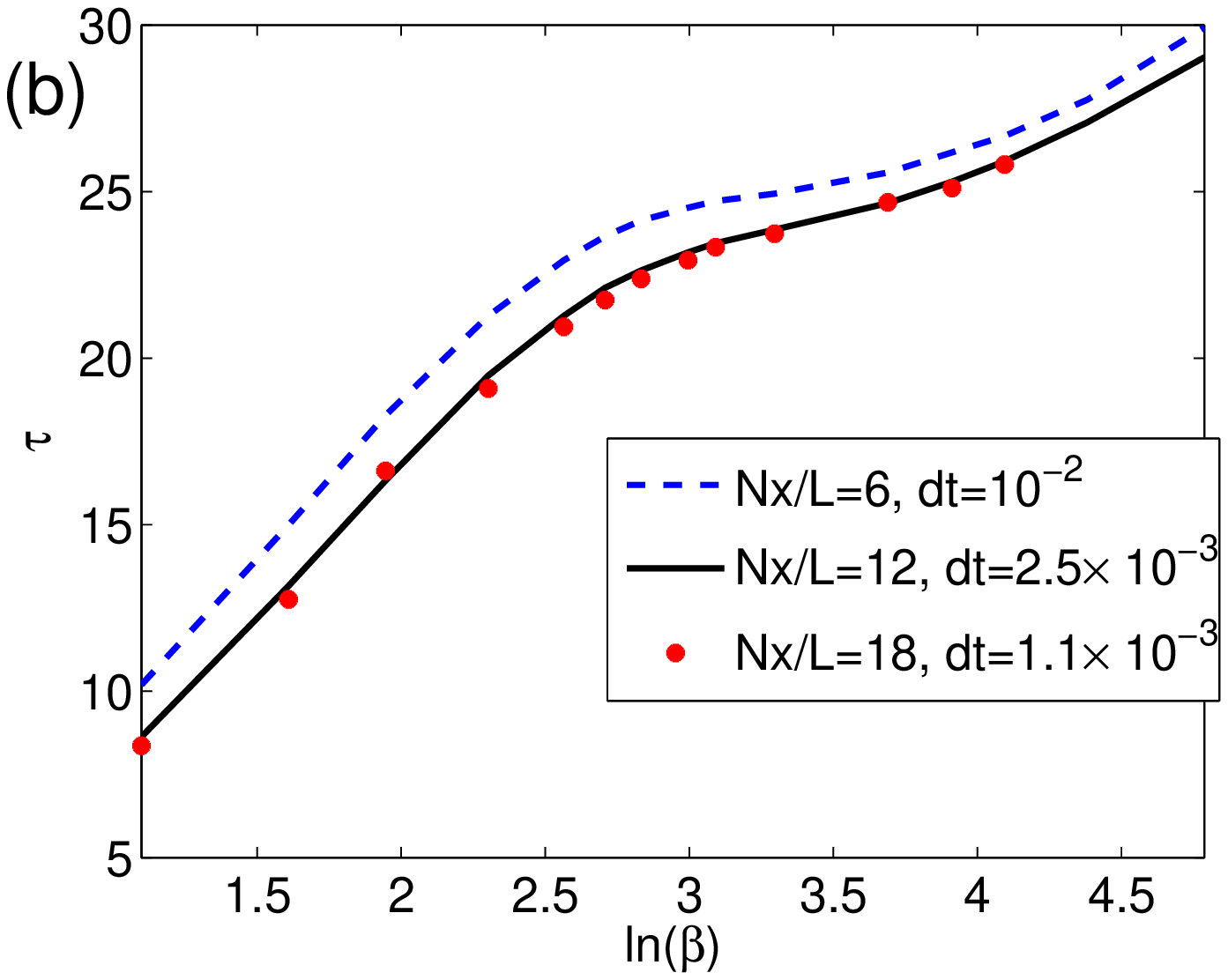}
\includegraphics[width=6cm]{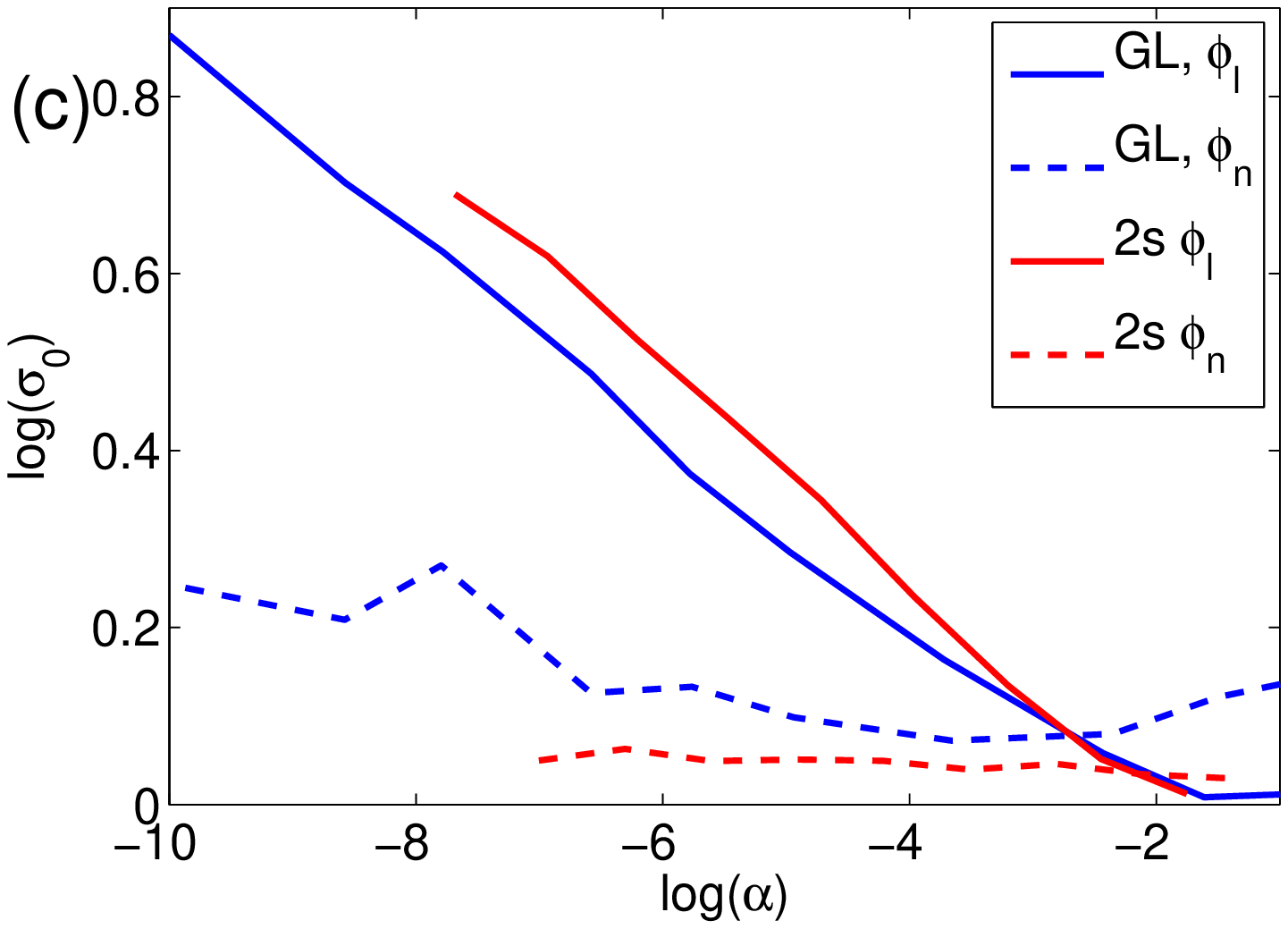}
}
\caption{(a): Logarithm of the escape probability as a function of $\beta$ for $L=7.07$,
$N=20~000$, $\Phi_n$ and three different values of $dt$ and $dx$.
(b): Average duration of reactive trajectories, in the same setting,
as a function of $\ln(\beta)$ for the same three resolutions.
(c): Comparison of the rescaled variance $\sigma_0$ of the escape probability as
a function of the logarithm of the escape probability, for the two reaction coordinates
$\Phi_l$ and $\Phi_n$. Both Allen--Cahn equation and the two-saddle model (see Eq.(\ref{sad2})) are shown.}
\label{conv}
\end{figure}

Estimates of $\ln \alpha$ (Fig.\ref{conv}a) and the average duration $\tau_r$ of reactive
trajectories (Fig.\ref{conv}b) are obtained for the different parameter cases. The result is
that increasing the spatial resolution does not improve the accuracy of $\ln \alpha $ whereas
it improves the accuracy on the average duration $\tau_r$
of reactive trajectories (Fig. \ref{conv}b).
Nevertheless, the qualitative behavior of $\tau_r$ is robust to the parameter $dx$ and $dt$.
In this article, we have decided to keep the values $dx = \frac16$ and $dt = 10^{-2}$ in all
the simulations.

We now discuss the effects of the number $N$ of initial trajectories and the choice of the
reaction coordinate $\Phi$ in the quality of the results. In practice, for most choices of
$\Phi$, the results converge in the limit $N \to \infty$. However, the bias and
variance of the escape probability estimate can behave rather differently depending on $\beta$
and the choice of $\Phi$ \cite{jcp}. We look here at the relative variance
\begin{equation}\label{sig0}
\sigma_0 = \sigma \frac{\sqrt{N}}{\alpha \sqrt{-\ln \alpha}}
\end{equation}
as a function of $\beta$ and for the two reaction coordinates $\Phi_l$ and $\Phi_n$.
Here $\sigma$ is the variance of the estimate of $\alpha$ obtained by the algorithm.
Note that in dimension 1 or when $\Phi$ is the committor, we have $\sigma_0 = 1$ (\cite{cg07,STCO}).
In order to simplify a bit the discussion, we also perform an analysis on the following
system with two degrees of freedom.
\begin{equation}\label{sad2}
d(x,y) = -\nabla V dt + \sqrt{\frac{2}{\beta}} (dW_1,dW_2),
V(x,y) = \frac{x^4}{4}-\frac{x^2}{2} -0.3\left(\frac{y^4}{4}-\frac{y^2}{2} + x^2 y^2\right).
\end{equation}
The main interest of this reduced model is that it displays a similar behavior of $\sigma_0$
as the escape probability $\alpha$ becomes small.
The reaction coordinates for (\ref{sad2}) are $\Phi_l(x,y) = \frac{x+1}{2}$
and $\Phi_n(x,y) = \frac12 \sqrt{(x+1)^2+y^2/2}$. We used the same definitions of $\mathcal{A}$, $\mathcal{B}$ and $\mathcal{C}$ as equation.~(\ref{defabc}). Note that this constitutes a slight change in the definition of $\mathcal{A}$ and $\mathcal{B}$ compared to what was used in \cite{jcp}. Figure \ref{conv}c shows the behavior
of $\ln \alpha$ as a function  of $\beta$ for both Allen--Cahn equation and the reduced model.

For the linear reaction coordinate $\Phi_l$, the variance of the estimate distribution increases
with $\beta$ and has a bias going to zero when $N \to \infty$. This bias increases when $\beta
\to \infty$. For the Euclidian reaction coordinate $\Phi_n$, one obtains better results in
the small model with
a smaller variance and bias. These results appear to be generic among models \cite{jcp} and is
the subject of further theoretical investigations.
Based on these results, we perform all the Allen--Cahn simulations
using the reaction coordinate $\Phi_n$ in (\ref{GLobs}).
These are done with a rather small number of trajectories $N=1000$, when displaying
qualitative properties, and $N = 20~000$ for more quantitative results.

\section*{Acknowledgements}
The research leading to these results has received funding from the European Research Council under the European Union's seventh Framework Programme (FP7/2007-2013 Grant Agreement no. 616811) (F. Bouchet). The authors thank E. Vanden Eijnden for helpful suggestions on the literature on diffusive processes as well as M. Argentina for comments on Goldstone modes.

\end{document}